\documentclass[a4paper]{jpconf}
\usepackage{graphicx}

\newcommand{\qsq}{Q^{2}}
\newcommand{\he}{^{4}\mbox{He}}
\newcommand{\hy}{^{1}\mbox{H}}
\newcommand{\apv}{A_{\rm{LR}}}
\newcommand{\alr}{A_{\rm{LR}}}
\newcommand{\gevc}{\,\mbox{GeV}^2}
\newcommand{\sinW}{\sin^2\theta_{W}}
\newcommand{\ges}{G_{E}^{s}}
\newcommand{\gms}{G_{M}^{s}}

\newcommand{\Moller}{\mbox{ M{\o}ller}}
\newcommand{\gep}{G_{E}^{p}}

\newcommand{\gen}{G_{E}^{n}}

\begin{document}

\title{Strange Vector Form Factors from Parity-Violating Electron Scattering}

\author{Kent Paschke$^1$, Anthony Thomas$^2$, Robert Michaels$^3$, and David~Armstrong$^4$}

\address{$^1$ University of Virginia, Charlottesville, Virginia 22903, USA}
\address{$^2$ University of Adelaide, South Australia 5005, Australia}
\address{$^3$ Jefferson Lab, Newport News,  Virginia 23606, USA }
\address{$^4$ College of William and Mary, Williamsburg, Virginia 23187, USA}

\begin{abstract}
The simplest models might describe the nucleon as 3 light quarks, but this 
description would be incomplete without inclusion of the sea of glue and 
$q\bar{q}$ pairs which binds it. Early indications of a particularly large 
contribution from strange quarks in this sea to the spin and mass of the 
nucleon motivated an experimental program examining the role of these 
strange quarks in the nucleon vector form factors.  The strangeness 
form factors can be extracted from the well-studied electromagnetic 
structure of the nucleon using parity-violation in electron-nuclear 
scattering to isolate the effect of the weak interaction. With high 
luminosity and polarization, and a very stable beam due to its 
superconducting RF cavities, CEBAF at Jefferson Lab is a precision 
instrument uniquely well suited to the challenge of measurements of the 
small parity-violating asymmetries. The techniques and results of the two 
major Jefferson Lab experimental efforts in parity-violation studies, 
HAPPEX and G0, as well as efforts to describe the strange form factors in 
QCD, will be reviewed.
\end{abstract}

\section{Introduction}

The achievement which clearly identified Quantum Electrodynamics (QED) as 
the correct theory of the electromagnetic interaction was the quantitative 
explanation of the measurement of the Lamb shift in hydrogen. For the 
strong force we have Quantum Chromodynamics (QCD), a local gauge theory 
built on color, which has been verified with considerable precision in 
the high energy, ``asymptotically free'' regime. However, in the 
nonperturbative regime, where QCD is truly strong, it is still being
tested, as our capacity for calculation using lattice QCD grows.
By analogy with the Lamb shift in QED, a fundamental test of QCD in the 
nonperturbative regime is its capacity to accurately explain vacuum 
polarization. Since there are no strange valence quarks in the proton, 
the strangeness form factors of the nucleon present the ideal testing 
ground.

Historically, the enormous interest in the strangeness form factors arose 
from two things. The discovery of the EMC spin crisis, with its very small 
fraction of the proton spin carried by quarks, was widely interpreted in 
terms of a large negative fraction of the proton spin carried by strange 
quarks -- $\Delta s = -0.16 \pm 0.08$~\cite{Jaffe:1989jz}.
At about the same time, analysis of the octet baryon masses in terms of 
SU(3) symmetry suggested values of the strangeness sigma commutator,
$\sigma_s = \langle p | m_s \bar{s} s| p \rangle$, as large as 
330~MeV~\cite{Kaplan:1988}. Since $\sigma_s = 330$~MeV represents a 
contribution to the nucleon mass of more than one third from 
a single sea quark flavor, this was an astonishing result. Both of these 
indications suggested a far greater role for the strange quark in proton 
structure than had been anticipated before. Modern lattice estimates 
of the strange sigma commutator suggest a value almost an order of 
magnitude smaller~\cite{Young:2009zb}. In addition, it seems likely
that SU(3) breaking leads to a considerably smaller value of
$\Delta s$~\cite{Bass:2009ed}. Nevertheless, in the late 1980's the older 
values were a powerful motivation for the experimental program.

In this environment, a number of authors estimated the strange quark 
charge radius and the strange magnetic moment of the proton in various 
models. Indeed, almost the entire arsenal of hadronic models have been 
brought to bear on the question of the strange form factors, including, 
but not limited to, vector meson dominance, the Skyrme model, kaon loops, 
the chiral quark model, dispersion relations, the NJL model, the 
quark-meson coupling model, the chiral bag model, heavy-baryon chiral 
perturbation theory, {\em etc.}; reviews of these are available elsewhere 
\cite{BeckHolstein,KumarSouder}.
Many reported large values. For example, Jaffe \cite{Jaffe:1989} found 
$\langle r^2_s \rangle = 0.16 \pm 0.06$~fm$^2$ and $\mu_s = -0.31 \pm 0.09
\mu_N$ , while Jaffe and Manohar observed \cite{Jaffe:1989jz} that until 
``such time as these experiments are perfected, the prediction of 
$F_2^{(0)}$ remains an excellent goal for theorists who think they have 
understood the flavor structure of the proton.''

This excited theoretical discussion inspired novel experimental 
suggestions as to how one might actually measure these elusive quantities 
and it became clear that parity-violating electron scattering was the tool 
of choice \cite{McKeown,Beck}. Thus began a rigorous and demanding program 
lasting more than 20 years, which as we report, has experimentally defined 
the strange electric and magnetic form factors of the proton. This program 
would not have been possible without the long term support of funding 
agencies and major laboratories, including MIT Bates, Mainz and, of 
course, Jefferson Lab.

Ignoring the tremendous technological and experimental challenges, the 
principle of the measurement of the strange vector form factors is simple. 
There are three vector operators whose matrix elements must be determined, 
$\bar{q} \gamma^\mu q$ with $q=u, \, d$ and $s$. Using charge symmetry, 
which is respected at better than 1\% by the strong force, one obtains two 
constraints from the proton and neutron electromagnetic form factors. For 
the third constraint, the fact that the weak vector charges differ from 
the electromagnetic charges means that the measurement of parity-violating 
electron scattering from the proton is sufficient. Indeed, a measurement 
of the left-right asymmetry, $\alr$, in the scattering of longitudinally 
polarized electrons from an unpolarized proton target yields the strange 
form factors through
\begin{eqnarray}
\alr = -\frac{G_{F} \qsq}{4\pi\alpha\sqrt{2}} \left\{   (1-4\sinW) - \frac{\epsilon G^p_E G^n_E + \tau G^p_MG^n_M}{\epsilon (G^p_E)^2 + \tau (G^p_M)^2}   - \frac{\epsilon G^p_E G^s_E + \tau G^p_MG^s_M}{\epsilon (G^p_E)^2 + \tau (G^p_M)^2} \right. \nonumber \\      
\left.  - \frac{(1-4\sinW)\epsilon^{\prime} G^p_M G^p_A}{\epsilon (G^p_E)^2 + \tau (G^p_M)^2} \right\} .
\end{eqnarray}
Here $\tau = \qsq/4M_p^2$,
$\epsilon = \left(1 + 2(1+\tau)\tan^2\frac{\theta}{2}\right)^{-1}$, and 
$\epsilon^{\prime} = \sqrt{\tau(1+\tau)(1-\epsilon^2)}$.
$G_E^s(Q^2)$ and $G_M^s(Q^2)$ are the strange electric and strange 
magnetic form factors, respectively, and parameterize the strange quark 
contribution to the vector structure of the nucleon.
Also appearing in this expression is the axial form factor $G^p_A$, 
which becomes significant at backward scattering angles where 
$\epsilon^{\prime}$ is larger. To lowest order, this form factor is the 
same as is measured in charged-current neutrino scattering. However, for 
electron scattering, large radiative corrections are expected to modify 
$G^p_A$.
Decomposing $G^p_A$ into isoscalar $G^e_A(T=0)$, and $G^e_A(T=1)$ pieces, 
Zhu {\em et al.}~\cite{Zhu:2000gn} found, in a model-dependent analysis, 
that the radiative corrections for the isoscalar piece are small and 
under reasonable theoretical control, while the isovector part is large 
($\sim$~30\%) and less-well constrained theoretically. This latter term 
is interesting in its own right, as it includes the effective 
parity-violating coupling of the photon to the nucleon, the so-called 
``anapole'' term~\cite{anapole}, and, as we will describe, the 
parity-violating electron scattering program described here can also
shed some light on this topic.

In this chapter, we first discuss the prediction of strange form factors 
from modern theory, focusing on lattice QCD. We then describe the 
experimental aspects of this program at Jefferson Lab. Then we present 
briefly the global analysis which yields the strange form factors, 
compare the extracted form factors with theory, and provide an outlook 
for the future.

\section{Calculation of strange form factors within QCD}
While the theoretical study of strange form factors within QCD cannot lay 
claim to two decades of effort, it is not so far off. Indeed, the initial 
studies of the electromagnetic form factors of the octet baryons began in 
the early 1990's, with work by Leinweber, Woloshyn and 
Draper~\cite{Leinweber:1990dv}. It was these studies which gave direct 
information on the contribution to the baryon form factors from individual 
valence quarks that eventually led, some 14 years later, to remarkably 
accurate predictions~\cite{Leinweber:2004tc,Leinweber:2006ug} of the 
strange quark form factors. The technique used in this work was 
indirect~\cite{Leinweber:1999nf}, so that for the strange magnetic moment 
it was necessary to combine lattice QCD calculations of the valence form 
factors, under the assumption of charge symmetry, with experimental 
information on the octet magnetic moments. This work required control of 
the extrapolation of the valence moments as a function of quark mass in 
both quenched and full QCD~\cite{Young:2002cj}. Only in the last two years 
(so again two decades from the first calculations) has it been possible to 
make a direct calculation in full QCD, with the Kentucky group reporting 
very accurate values of the strange magnetic moment of the proton in very 
good agreement with the earlier indirect calculations and with experiment.

\subsection{Indirect method}
\label{sec:pvestheory:indirect}
\begin{figure}[tbp]\begin{center}
{\includegraphics[height=5.3cm,angle=90]{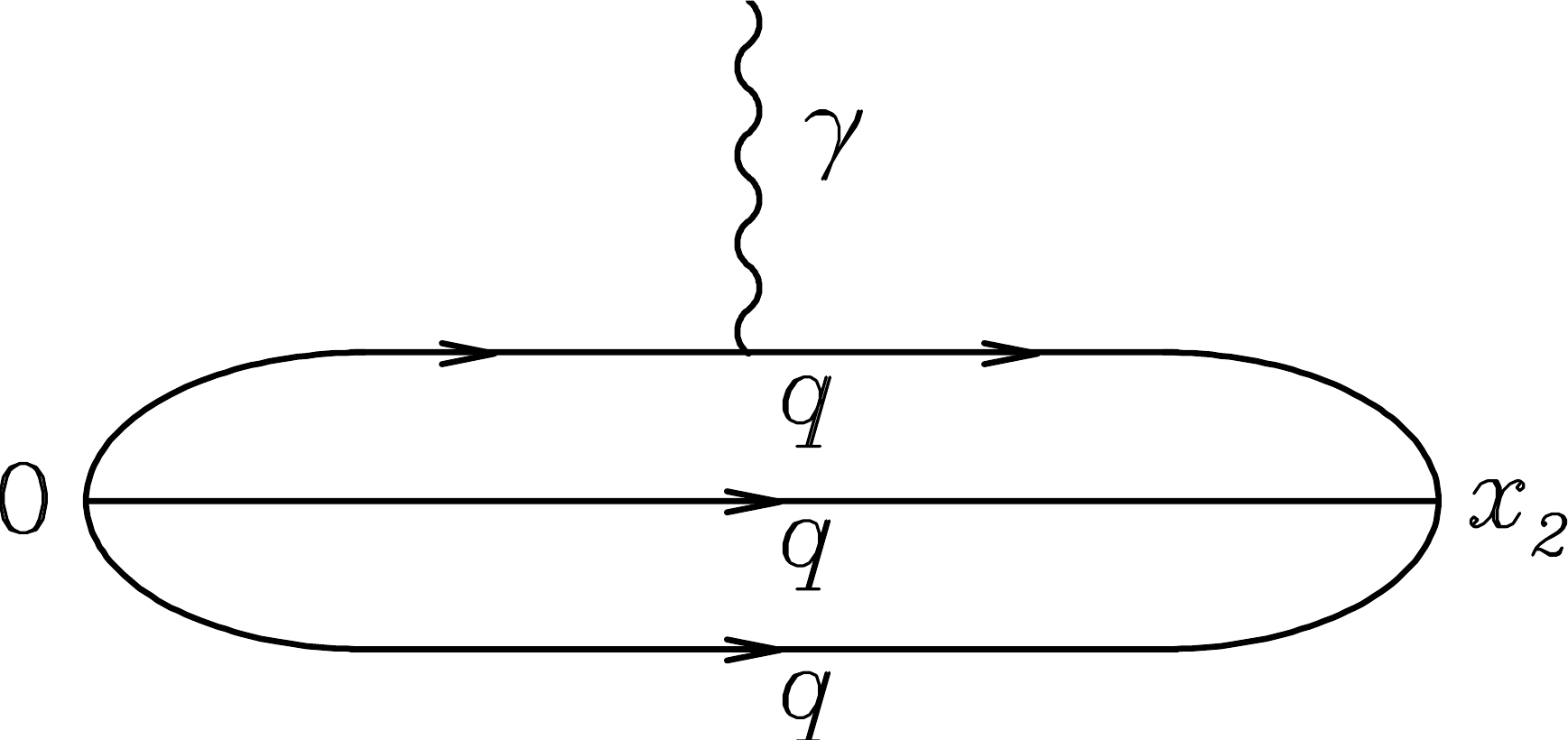} \hspace{0.8cm}
 \includegraphics[height=5.3cm,angle=90]{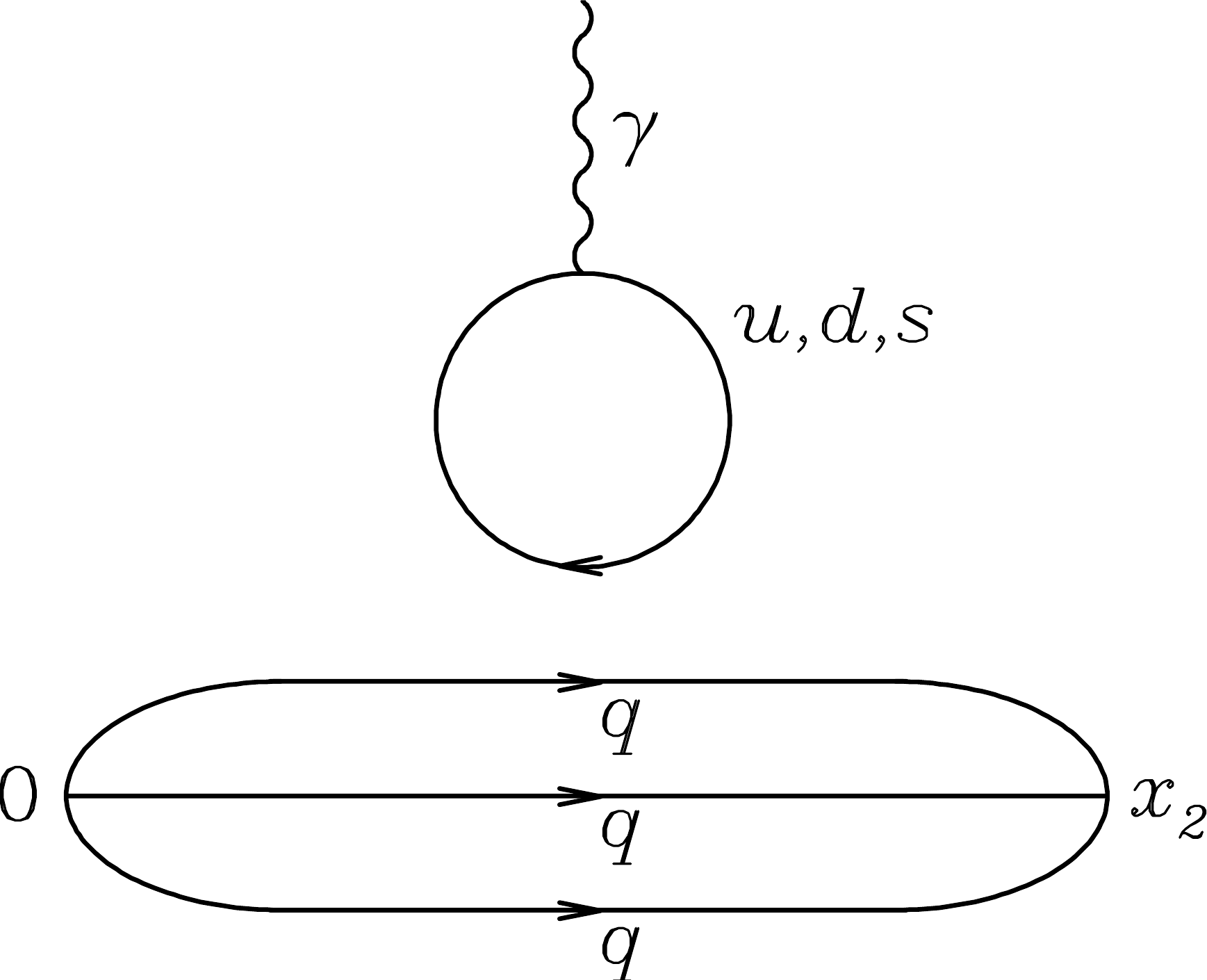}}
\caption{Diagrams illustrating the two topologically different
insertions of the current within the framework of lattice QCD.
}
\label{topology}
\vspace{-0.5cm}
\end{center}
\end{figure}

As illustrated in Fig.~\ref{topology}, the three point function required
to extract a magnetic moment in lattice QCD involves two topologically
distinct processes. (Of course, in full QCD these diagrams incorporate
an arbitrary number of gluons and quark loops.)
The left-hand diagram illustrates the connected insertion of the current 
to one of the ``valence'' quarks of the baryon.  In the right-hand 
diagram the external field couples to a quark loop. The latter process, 
where the loop involves an $s$-quark, is entirely responsible for the 
strange quark contribution to the nucleon form factor.

Under the assumption of charge symmetry~\cite{ChargeSymm},
the magnetic moments of the octet baryons
satisfy~\cite{Leinweber:1999nf}:
\begin{eqnarray}
p &=& e_u\, u^p + e_d\, d^p + O_N  \, \, ; \, \,
n = e_d\, u^p + e_u\, d^p + O_N  \, , \nonumber \\
\Sigma^+ &=& e_u\, u^{\Sigma} + e_s\, s^\Sigma + O_\Sigma  \, \, ; \, \,
\Sigma^- = e_d\, u^{\Sigma} + e_s\, s^\Sigma + O_\Sigma  \, , \nonumber  \\
\Xi^0 &=& e_s\, s^\Xi + e_u\, u^{\Xi} + O_\Xi  \, \, ; \, \,
\Xi^- = e_s\, s^\Xi + e_d\, u^{\Xi} + O_\Xi  \, . \nonumber  \\
\label{equalities}
\end{eqnarray}
Here, $p$ and $\Xi^-$ are the physical magnetic moments of the proton 
and $\Xi^-$, and similarly for the other baryons. The valence $u$-quark 
sector magnetic moment in the proton, corresponding to the left hand side of 
Fig.~\ref{topology}, is denoted $u^p$. Charge symmetry has been used to 
replace the $d$-quark contribution in the neutron by $u^p$, $d$ in the 
$\Sigma^-$ by $u$ in the $\Sigma^+$ ( $u^\Sigma$), and so on. The labels 
on quark magnetic moments allow for the environment sensitivity implicit 
in the three-point function. That is, the naive expectations of the 
constituent quark model, namely $u^p/u^{\Sigma} = u^n/u^{\Xi} = 1$, 
may not be satisfied.
The total contribution from quark-loops, $O_N$, contains sea-quark-loop 
contributions (right hand side of Fig.~\ref{topology}) from $u$, $d$ and $s$ quarks. 
By definition
\begin{eqnarray}
O_N &=& \frac{2}{3} \,{}^{\ell}G_M^u - \frac{1}{3} \,{}^{\ell}G_M^d -
\frac{1}{3} \,{}^{\ell}G_M^s \, , \\
&=& \frac{{}^{\ell}G_M^s}{3} \left ( \frac{1 -
{}^{\ell}R_d^s}{{}^{\ell}R_d^s } \right ) \, ,
\label{OGMs}
\end{eqnarray}
where the ratio of $s$- and $d$-quark {\it loops},
${}^{\ell}R_d^s \equiv {{}^{\ell}G_M^s}/{{}^{\ell}G_M^d}$, is
expected to lie in the range (0,1).
Note that, in deriving Eq.~(\ref{OGMs}), we have used charge symmetry 
to set ${}^{\ell}G_M^u = {}^{\ell}G_M^d$.
Since the chiral coefficients for the $d$ and $s$ loops in the right hand side 
of Fig.~\ref{topology} are identical, the main difference comes from the mass of the
$K$ compared with that of the $\pi$.

With a little algebra, $O_N$, and hence $G_M^s (\equiv {{}^{\ell}G_M^s})$,
may be isolated from Eqs.~(\ref{equalities}) and (\ref{OGMs}):
\begin{eqnarray}
G_M^s &=& \left ( {\,{}^{\ell}R_d^s \over 1 - \,{}^{\ell}R_d^s }
\right ) \left [ 2 p + n - {u^p \over u^{\Sigma}} \left ( \Sigma^+ -
\Sigma^- \right ) \right ] ,
\label{GMsSigma} \\
G_M^s &=& \left ( {\,{}^{\ell}R_d^s \over
1 - \,{}^{\ell}R_d^s } \right ) \left [
p + 2n - {u^n \over u^{\Xi}} \left ( \Xi^0 - \Xi^- \right )
 \right ] .
\label{GMsXi}
\end{eqnarray}
After incorporating the experimentally measured baryon
moments \cite{PDG} (in nuclear magnetons, $\mu_N$),
Eqs.~(\ref{GMsSigma}) and (\ref{GMsXi}) become:
\begin{eqnarray}
G_M^s &=& \left ( {\,{}^{\ell}R_d^s \over 1 - \,{}^{\ell}R_d^s } \right ) 
\left [
3.673 - {u^p \over u^{\Sigma}} \left ( 3.618 \right ) \right ] ,
\label{ok} \\
G_M^s &=& \left ( {\,{}^{\ell}R_d^s \over
1 - \,{}^{\ell}R_d^s } \right ) \left [
-1.033 - {u^n \over u^{\Xi}} \left ( -0.599 \right ) \right ] .
\label{great}
\end{eqnarray}
We stress that these expressions for $G_M^s$ are exact consequences
of QCD, under the assumption of charge symmetry.

Equating (\ref{ok}) and (\ref{great}) provides a linear relationship
between $u^p/u^{\Sigma}$ and $u^n/u^{\Xi}$ which must be satisfied
within QCD under the assumption of charge symmetry.
Clearly this linear relationship is inconsistent with the assumption 
of universality of the valence quark moments ({\it i.e.}, the point 
(1.0,1.0) does not lie on the empirical line). In addition, it is 
important from the point of view of reducing systematic errors that the 
linear relation involves only {\em ratios} of moments, which can be 
calculated more accurately than absolute values.
\begin{figure}[tbp]\begin{center}
{\includegraphics[height=7.8cm,angle=90]{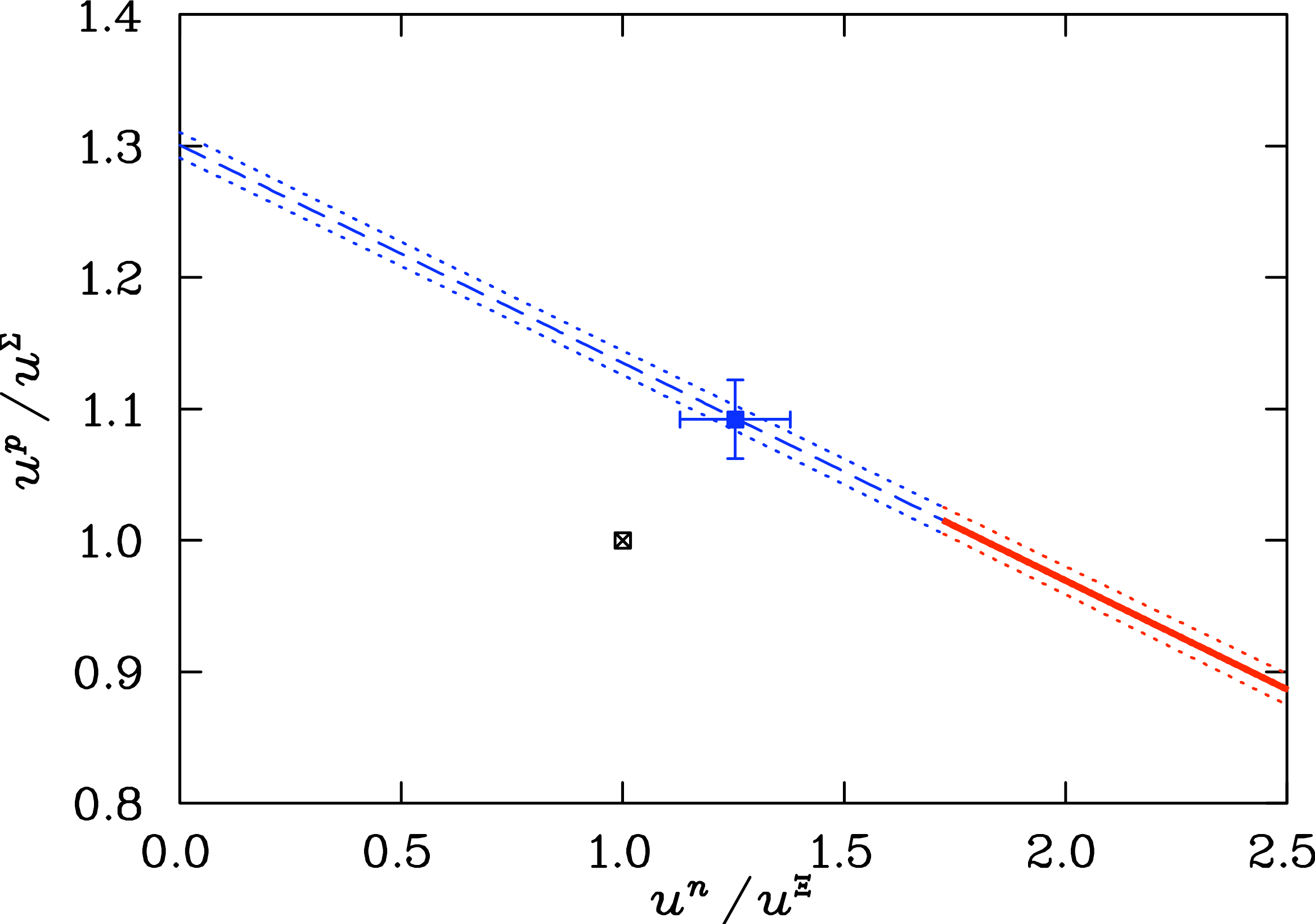}}
\vspace*{-0.5cm}
\caption{The constraint (dashed $G_M^s(0) < 0$, solid $G_M^s(0) > 0$) on 
the ratios $u^p/u^{\Sigma}$ and $u^n/u^{\Xi}$ implied by charge symmetry 
and experimental moments.  Experimental uncertainties are indicated by the 
dotted bounds.  The assumption of environment-independent quark moments is 
indicated by the crossed square.  The final result (chiral-corrected 
extrapolation of lattice results) is illustrated by the filled square, 
which does indeed lie on the charge symmetry line.}
\label{SelfCons}
\vspace{-0.5cm}
\end{center}\end{figure}

The actual lattice simulations gave the individual valence quark moments 
as a function of pion mass in quenched QCD. These were then fit using 
finite range regulated effective field theory on a finite volume, with 
coefficients for quenched QCD, which were then replaced by full QCD chiral 
corrections on an infinite volume. From this chiral extrapolation 
procedure, the ratios of the valence (connected) $u$-quark contributions, 
${u^p}/{u^\Sigma}$ and ${u^n}/{u^\Xi}$ were calculated.
The final results
\begin{equation}
\frac{u^p}{u^\Sigma} = 1.092\pm 0.030
\quad
\mbox{and}
\quad
\frac{u^n}{u^\Xi} = 1.254\pm 0.124
\label{uRatios}
\end{equation}
are plotted in Fig.~\ref{SelfCons}.
The precision of these results follows from the use of correlated
ratios of moments which act to reduce uncertainties associated with
the lattice spacing, the regulator mass and statistical fluctuations
\cite{Leinweber:2005bz}.
This result leaves no doubt that $G_M^s$ is negative.  The fact that
this point lies exactly on the constraint curve is highly nontrivial,
and provides a robust check of the validity of the analysis techniques.

While $G_M^s$ was certainly negative, it remained to set the
magnitude.  This required an estimate of the strange to light
sea-quark loop contributions, ${}^{\ell}R_d^s$.  Earlier estimates of
$^{\ell}R_d^s$ had been based on the constituent quark model.  A more
reliable approach is to estimate the loops using the same successful
model invoked to correct the quenched results to full QCD
\cite{Young:2002cj,Young:2004tb}.
Allowing the dipole mass parameter to vary between 0.6 and 1.0 GeV
provides ${}^{\ell}R_d^s = {G_M^s}/{G_M^d}=0.139 \pm 0.042$.
A complete analysis of the errors associated with the determination of
$G_M^s$ using Eqs.~(\ref{GMsSigma}), (\ref{GMsXi}) and (\ref{uRatios})
is reported in Ref.~\cite{Leinweber:2005bz}.  The uncertainty is dominated
by the statistical errors included in Eq.~(\ref{uRatios}) and the
uncertainty just noted for ${}^{\ell}R_d^s$.  The final result for the
strangeness magnetic moment of the nucleon is
\begin{equation}
{G_M^s} = -0.046 \pm 0.019\ \mu_N \, .
\label{GMs}
\end{equation}

The calculation of the strange quark charge radius involved the same ideas 
presented in Eqs.~(\ref{ok}) and (\ref{great}) but in this case the octet 
data are not available, so one could not use ratios of valence properties 
to reduce errors. Nevertheless the result,
$\langle r^2 \rangle^p_s = 0.001 \pm 0.004 \pm 0.004$~fm$^2$, was 
remarkably precise. Finally, we note that this technique has also been 
used to extract the strange magnetic form factor at finite 
$Q^2$~\cite{Wang:1900ta}.

\subsection{Direct method}
The direct calculation of the disconnected strange quark loop has proven 
extremely difficult in lattice QCD, which is why the indirect techniques 
described above were applied first. The first direct calculations 
\cite{Dong,Lewis} were unable to extract robust signals for the strange 
form factors. However, just in the last two years, the Kentucky group has 
succeeded in beating down the noise to obtain a very convincing signal for 
the magnetic and electric strange quark form factors of the proton as a 
function of $Q^2$~\cite{Doi:2009sq}. For the present time, the results 
have been reported at somewhat heavy light quark masses -- corresponding 
to pion masses around 500~MeV. Nevertheless the strange
magnetic moment found by Doi {\it et al.},
namely $\mu_s = -(0.017 \pm 0.025 \pm 0.007) \mu_N$,
is in excellent agreement with the values quoted above. 
Indeed, a simple estimate of the effect of moving to the physical 
light quark mass would make the agreement even better. The result for 
the strange electric charge radius was also in agreement with the value 
quoted above.

In conclusion, the theoretical status of the calculation of the strange 
electric and magnetic form factors seems to be very sound, with excellent 
agreement between the different techniques employed.

\section{Experimental overview}

As described above, the strange vector form factors are accessible 
through the precision measurement of the helicity-correlated cross 
section asymmetry in the elastic scattering of polarized electrons 
from an unpolarized target, $\apv$.
This asymmetry is small, on the order of 1-100~parts~per~million (ppm) 
for the kinematics which are typically of interest, $\qsq<1~\gevc$,
and must be measured with a precision in the range of 10\% or better.  
Experiments of this nature are optimized to the challenges of precision 
measurement of very small asymmetries, which require large high count 
rates and low noise to achieve statistical precision as well as a careful 
regard for potential systematic errors.

One common feature of all measurements of parity-violation in electron 
scattering is a rapid flipping of the electron beam helicity, allowing a 
differential measurement between opposing polarization states on a short 
timescale.  The enabling technology for these measurements lies in the 
semiconductor photo-emission polarized electron source, which allows rapid 
reversal of the electron polarization while providing high luminosity, 
high polarization, and a high degree of uniformity between the two beam 
helicity states.  Developments with the polarized source at Jefferson Lab 
critical to the success of this program are described elsewhere in this
volume \cite{BSM}.

The following sections will describe a series of measurements in 
parity-violating in electron-nucleus scattering which took place over the 
first decade of the operation of Jefferson Lab: the HAPPEX measurements 
in Hall~A and the G0 measurements in Hall~C.

\section{HAPPEX-I}

\begin{figure}\begin{center}
\includegraphics[height=1.6in,angle=0]{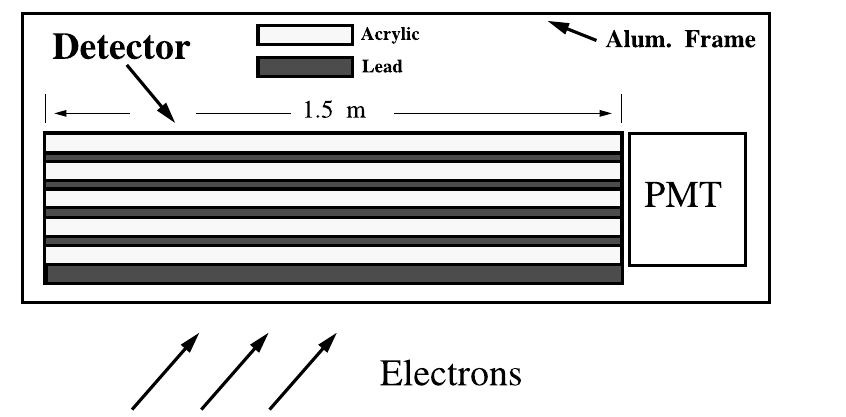}
\caption{Schematic of the focal plane detector used by HAPPEX-I.
The scattered electrons strike a lead-acrylic shower counter whose
light is collected by a PMT and integrated over a helicity period.}
\label{fig:happex1_detector}
\end{center}\end{figure}

The Hall~A Proton Parity Experiment (HAPPEX) ran in 1998 and 1999
and pioneered parity-violation study at Jefferson Lab 
\cite{happex1_aniol1,happex1_aniol2,happex1_prc}.
HAPPEX measured $\apv$ in elastic electron-proton scattering at the 
kinematics $\langle \theta_{\rm lab} \rangle = 12.{3^\circ}$
and $\langle Q^2 \rangle = 0.477$ (GeV/c)$^2$.  
The Hall~A High resolution spectrometers (HRS)~\cite{A-NIM} were used to 
focus the scattered electrons onto a total absorption shower counter in 
the spectrometer focal plane (Fig.~\ref{fig:happex1_detector}).
The HRS cleanly isolated the elastic scattered electrons, suppressing
background from inelastics or low-energy secondaries. The detector in
each spectrometer was instrumented with a single photomultiplier (PMT)
tube. Rather than counting individual pulses, the anode current from
this PMT was simply integrated over the 33~millisecond helicity
periods, and the result combined with that from a consecutive window
of opposite electron-beam helicity, to form the asymmetry measurement
``pair''.  With a detected rate of 2~MHz, the scattering rate asymmetry
was measured at 15~Hz with a precision of 0.38\%.  The selected
kinematics corresponded to the smallest angle and largest energy
possible with the Hall~A HRS spectrometers, which maximized the figure
of merit for a first measurement.

HAPPEX-I stimulated improvements in both polarized
source technology and polarimetry at Jefferson Lab.
In the 1998 run the experiment used a $I = 100 \;~\mu$A beam 
with $P = 38$\% polarization produced from a bulk GaAs crystal,
while in the 1999 run HAPPEX-I became the first experiment to
use a strained GaAs photocathode to measure a parity-violating 
asymmetry in fixed-target electron scattering.
This improved the figure-of-merit $P^2 I$ with 
$P$=70\% and $I$=35$~\mu$A but also 
required several refinements of the techniques used to control
systematic errors associated with the laser at the
polarized source, in addition to the usual
feedback control of the helicity-correlated charge asymmetry.
The active layer of the strained photocathode was a thin ($\sim$100 nm) layer of GaAs 
grown on GaAsP. The mismatch between the two lattices produced
a strain in the GaS that breaks an energy-level degeneracy, 
allowing selective photoexcitation of a specific polarization state and a 
theoretical maximum 100\% electron 
polarization \cite{happex1_NAK91,happex1_MAR92}. During HAPPEX-I, a 70\% polarization 
was achieved, and at the present time 85\% is available due to
further refinements in the technology.
The strain, however, introduced an anisotropy in the quantum
efficiency of the cathode, making it the dominant source
of analyzing power in the system. 
A rotatable half-wave plate inserted downstream of the Pockels
cell provided the ability to rotate the laser beam's polarization
ellipse, which reduced the sensitivity to the analyzing power,
as determined by scans of the helicity-correlated position differences and
charge asymmetries versus the angle of this half-wave plate 
\cite{happex1_prc}.  A half-wave plate upstream of the Pockels cell was 
aligned to invert the sign of incident linear polarization, which toggled 
the sign of the laser circular polarization, and therefore the electron 
beam helicity, relative to the Pockels cell voltage settings.  The state 
of this waveplate was toggled every 24--48 hours. This method of slow 
helicity reversal provided a way to cancel out some sources of possible 
systematic errors, including false asymmetries from electronics pickup and
certain helicity-correlated beam asymmetries.

During HAPPEX-I the Hall~A Compton polarimeter~\cite{Baylac} was 
commissioned and provided, for the first time, a continuous monitoring of 
the electron beam polarization with a run-to-run relative error of less 
than 2\%, and a total error on absolute polarization averaged over the run 
of 3.3\%. The dominant systematic uncertainty for this polarimeter lay in 
the corrections of the theoretical analyzing power for realistic detector 
performance. A second polarimeter was also used in Hall~A, based on 
M{\o}ller scattering from a polarized ferromagnetic foil~\cite{A-NIM}.  
The M{\o}ller polarimetery results carried a total uncertainty of 3.2\%, 
primarily due to uncertainty in the polarization of the electrons in the 
target foil. This polarimeter was invasive and could not be used as a 
continuous monitor, but results from measurements interspersed with 
data-taking were in good agreement with the results from the Compton 
polarimeter. These results agree also with measurements by a third 
polarimeter, based on Mott scattering in the 5~MeV region of the injector, 
which was quoted with a total error of about 7\%.

\begin{figure}[pbth]\begin{center}
\includegraphics[width=3.5in]{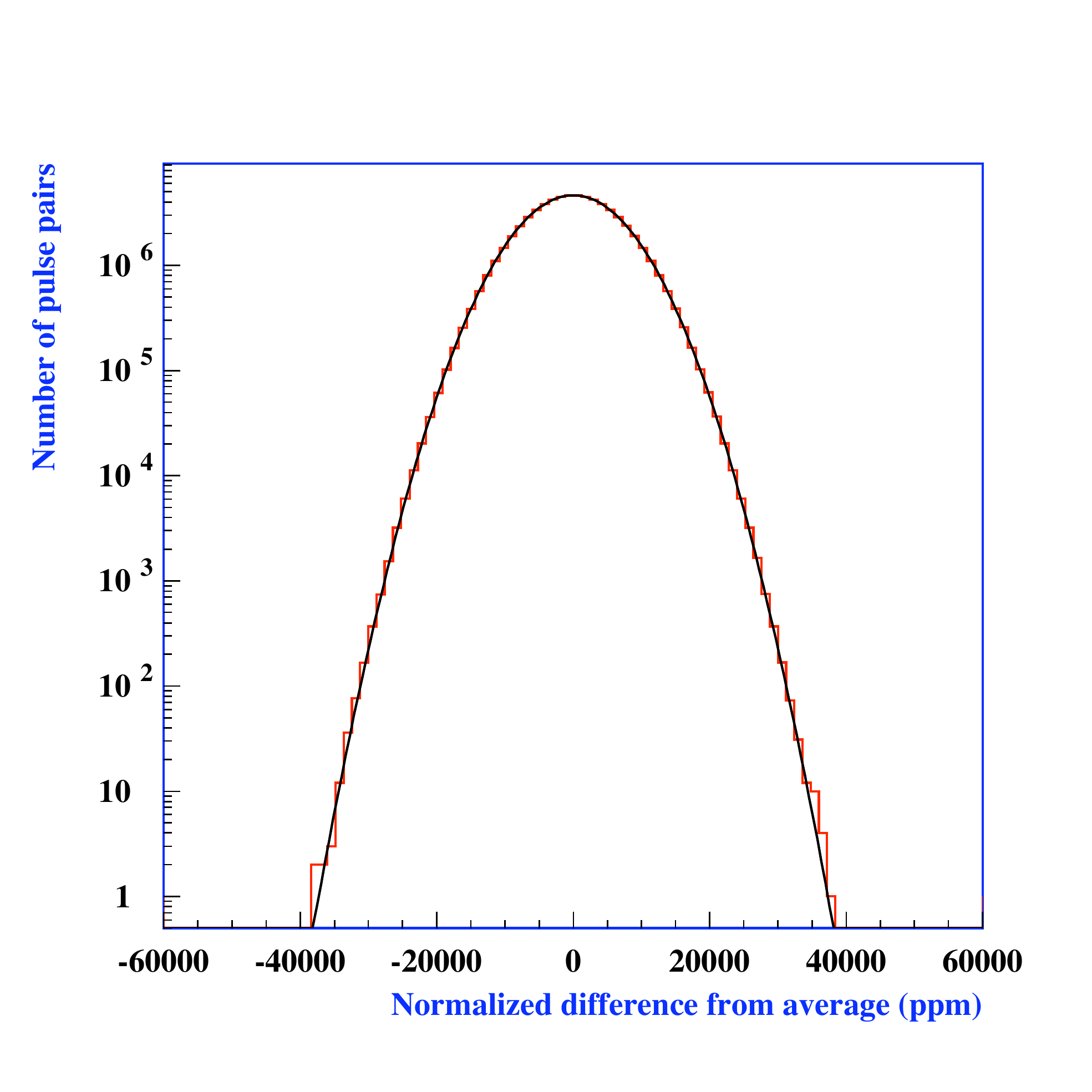}
\caption{The window pair asymmetries in ppm for HAPPEX-I,
normalized by the square root of beam intensity, with the
mean value subtracted off.  The curve is a Gaussian fit.}
\label{fig:happex1_pairasymplot}
\end{center}\end{figure}

\begin{figure}[phtb]\begin{center}
\includegraphics[width=2.6in,angle=90]{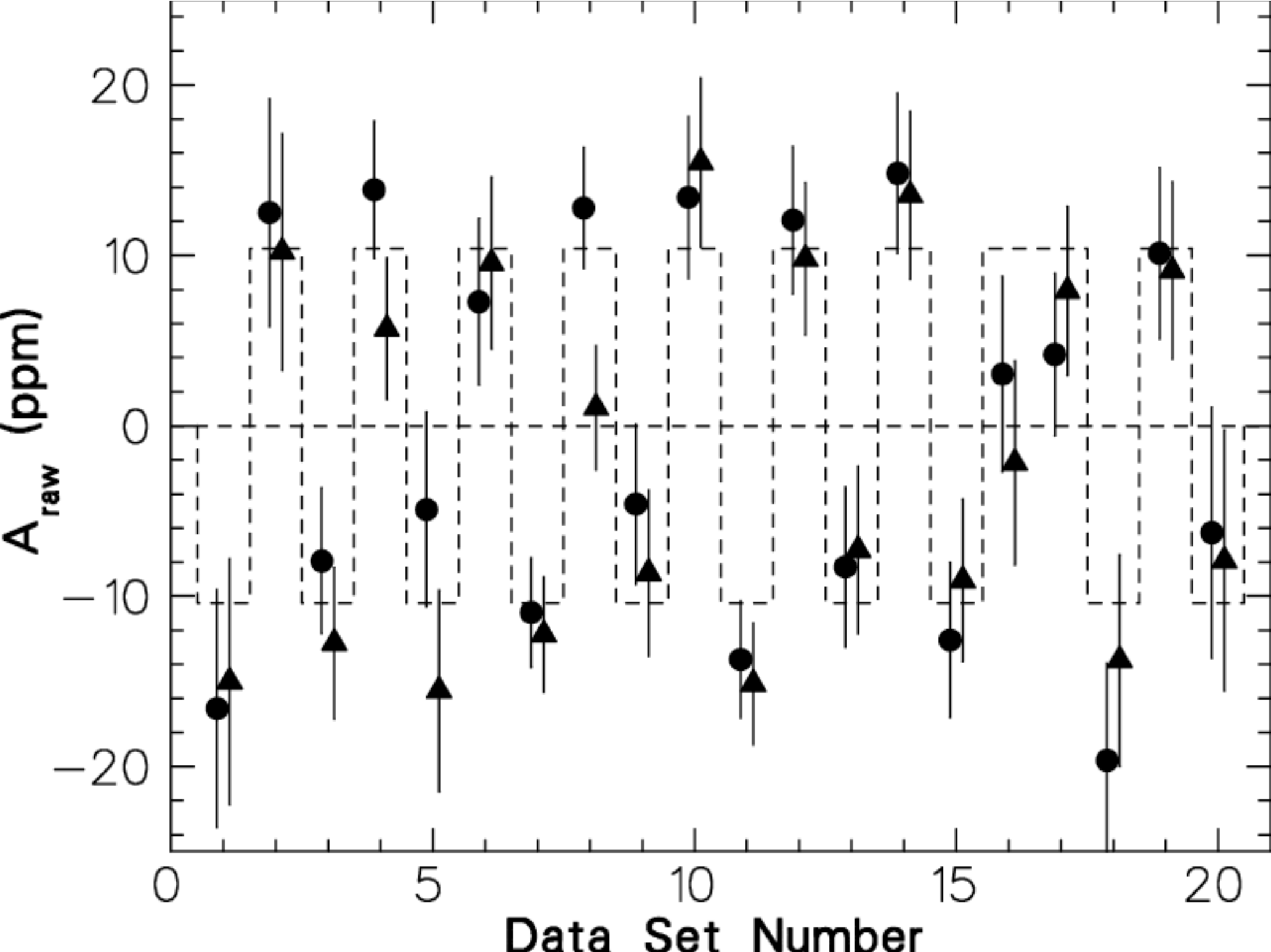}
\caption{Raw asymmetries (ppm) for HAPPEX-I in the 1999 run.
Each data set is $\sim$1 day of running.  The circles are for the left
spectrometer, triangles for the right spectrometer.  The step pattern
represents the insertion and removal of the half-wave plate which
reverses the sign.}
\label{fig:happex1_halfwave}
\end{center}\end{figure}

HAPPEX-I yielded a very clean physics result and was published
within a year after the experiment was completed.  Due to 
the high-quality of the Jefferson Lab polarized source and 
superconducting accelerator, the systematic errors associated
with the beam were found to be negligible compared to the statistical error.  
The accuracy
of the result was sufficient to rule out several theoretical
estimates of strangeness effects at moderately high $Q^2$ where 
it was thought the effects might have been large.
The distribution of window-pair asymmetries 
are shown in Fig.~\ref{fig:happex1_pairasymplot}; the distribution 
is Gaussian over 6 orders of magnitude, with a width consistent
with counting statistics.
Uncorrected asymmetries from the 1999 running period are shown
in Fig.~\ref{fig:happex1_halfwave}, averaged over the 24-48 hour periods
between half-wave plate insertions.  The measured asymmetry flips
sign cleanly with the insertable half-wave plate.  
After all corrections, the HAPPEX-I physics asymmetry was found to be 
\begin{eqnarray}\label{happex1_asy}
A = -15.05 \pm 0.98 \pm 0.56 \; {\rm ppm}, 
\end{eqnarray}
where the first error is statistical and the second error is
systematic. This latter includes the errors in the beam polarization,
background subtraction, helicity-correlated beam properties, and
$Q^2$.
Using this result, along with the calculated $G^{p}_A$ ~\cite{Zhu:2000gn} and the known 
values of the proton and neutron form factors, the experiment determined the
linear combination of strange
form factors $G_E^s + \eta G_M^s$ where $\eta = \tau G_M^{\gamma p}
/ \epsilon G_E^{\gamma p} = 0.392$ at the given kinematics.  
Thus,
\begin{equation} 
\label{eq:happex1_strff}
G_E^s + \eta G_M^s = 0.014 \pm 0.020 \pm 0.010,
\end{equation}
where the first error is the total experimental error
(statistical and systematic errors added
in quadrature) and the second error
is the error due to the 
``ordinary'' electromagnetic form factors,
which was dominated by the uncertainty in $G_M^n$.

The clean and rapidly published result of HAPPEX-I, encouraged the 
development of further, more accurate and ambitious parity-violation 
measurements at Jefferson Lab.

\section{The second generation HAPPEX experiments}

The second generation HAPPEX experiments ran in 
2004~\cite{Aniol:2005zf,Aniol:2005zg} and 2005~\cite{Acha:2006my} .  
Similar to the original HAPPEX experiment in technique, they measured 
forward-angle elastic scattering from the proton and from the $\he$ 
nucleus using the HRS to cleanly separate the elastic events.  A smaller 
scattering angle was achieved, about 6$^{\circ}$, by using a 
superconducting septum dipole magnet to bend the small scattering angles 
into the $12.5^{\circ}$ minimum acceptance for the HRS.  The figure of 
merit was again optimized at the maximum momentum accepted by the HRS, 
with a $\qsq\sim0.1\gevc$.
%

Elastic scattering from an isoscaler $0^+$ nucleus does not allow 
contributions from magnetic or axial-vector currents. At tree level, the 
parity-violating asymmetry for scattering from $\he$ can be written as
\begin{equation}
A_{PV}^{\rm He} = \frac{G_F \qsq}{4\pi\alpha\sqrt{2}}\left(  4\sinW 
+\frac{2\ges}{\gep+\gen}\right).
\end{equation}
With sufficient precision, measurement of this asymmetry isolates the 
strange electric form factor $\ges$ alone. For $\he$, nuclear model 
corrections to the asymmetry due to isospin mixing
\cite{Ramavataram:1994,Viviani:2007}, $D$-state admixtures 
\cite{Musolf:1993}, and meson-exchange contributions \cite{Musolf:1994} 
are all negligible at low $\qsq$.
In addition to the independence from nuclear models, $\he$ provided a 
particular advantage of a deeply-bound ground state, which separated the 
nearest inelastic level by about 20~MeV and allowed the HRS to easily 
isolate elastic scattering. The strong $\qsq$ dependence of the Mott 
cross section supports the figure-of-merit in the region 
$\qsq\sim0.1\gevc$.  Measurements at significantly higher $\qsq$ would be 
problematic, both due to figure-of-merit and control of the nuclear 
corrections.

The measurements used 20~cm long cryogenic targets, with $\sim 3$~GeV 
electron beam at currents from 35 to 55~$\mu$A.  The targets were a novel 
design, using a transverse flow of cryogen across the beam axis to avoid 
localized heating in regions of low fluid flow velocity.  A 
superconducting septum magnet captured scattered flux from around 
$6^{\circ}$ into the HRS acceptance, where the elastically scattered 
electrons in each spectrometer were focused onto a brass/quartz Cerenkov 
counter.  The signal from each counter was integrated over periods of 
about 33~milliseconds, resulting in an asymmetry measurements made at a 
rate 15~Hz with a precision of 540~ppm for the $\hy$ target and 1130~ppm 
for the $\he$ target.

The high precision of these measurements required careful control of 
helicity-correlated beam asymmetries. Detailed studies of the laser optics 
of the polarized source led to improved characterization of the optical 
elements, as well as an improved algorithm for aligning the electro-optic 
Pockels cell which is used to create the fast helicity flip.  Care was 
also taken to avoid magnifying any helicity-correlated orbit changes 
during beam acceleration and transport.  As a result, the 
helicity-correlated position differences, averaged over the course of
the run, was held to less than 2~nm for the Hydrogen measurement.

For the Helium measurement, electronics meant to drive additional 
helicity-correlated feedback systems created an unanticipated problem.  
An electrical control signal, indicating the beam polarization state to a 
hardware driver on the source laser table, interacted with devices on the 
electron beam line to steer the electron beam.  This led to large position 
differences, as large as 600~nm at the target, which were in fact related 
only to the electrical signal which indicated helicity, and not to the 
actual helicity of the electron beam. This was proven by the observation 
that the differences existed even when the electron beam was left 
unpolarized by switching off high-voltage to the Pockels cell used to 
create the laser polarization state. Changing the polarity of this control 
signal input to the suspect system was also seen to toggle the sign of the 
steering effect.  Careful checks revealed no electrical contamination of 
the data acquisition in the experimental hall; this effect was confined to 
the observed steering in the injector.  Before the hardware driver was 
switched off to remove this effect, additional production data was first 
collected with the control signal flipped to provide a degree of 
cancellation for effects of this electrical leakage. Because this effect 
related to the control signal and not to laser or electron beam 
polarization, it was also seen to cancel well when averaged over the 
half-wave plate helicity reversal.

In addition to controlling the beam asymmetries, a correction for measured 
helicity-correlated beam asymmetries was applied.  The sensitivities, that 
is, the changes in detected rate with changing beam parameters, were 
calibrated using periodic, deliberate beam modulation. Air core magnets in 
the arc leading to the hall moved the beam in a step-wise, sawtooth 
pattern, spanning the space of both position and angle at the target.  An 
energy vernier was used to modulate the energy as well.  Measuring the 
response of beam monitors and the detector to these individual modulations 
allowed for a measurement of the slopes which enforced the orthogonality 
of the motions, and avoided systematic misinterpretations of the 
sensitivities due to correlations in the beam motion or electronics noise.  
The results were consistent with expectations from simulation.  The small 
corrections for helicity-correlated beam asymmetries, and the estimated 
systematic uncertainty in those corrections, were based on the 
sensitivities measured using this beam modulation.

The detector, which relied on Cerenkov light generated by high-energy 
leptons in an electromagnetic shower, preferentially collected light from 
the direction of primary tracks in the spectrometer.  This directional 
sensitivity and low sensitivity to soft backgrounds, along with the 
detector location in the heavily shielded detector hut of the HRS, 
provided a very low background measurement.  The largest background in the 
measurement, due to quasi-elastic scattering from the aluminum windows of 
the target, was measured to be $0.8\pm0.25$\% of the signal from $^1$H and 
$1.8\pm0.2$\% for $\he$. This fraction was directly bounded by 
measurements of the production target with varying cryogenic gas 
densities.  The uncertainty in the background rate and in the asymmetry of 
the background (due to possible inelastic contributions which would not be 
detectable at very low fractional rate) combined for a total uncertainty 
of $1.3\%$ for $^1$H and $0.6\%$ for $\he$. Spectra of the high-resolution 
kinematic reconstruction in the HRS gave a range of $0.15\pm0.15\%$ for 
the quasielastic and inelastic fractional contributions to the $\he$ 
signal.  Additional backgrounds due to rescattering of inelastics in the 
spectrometer were estimated using dedicated calibration runs to measure 
rescattering probabilities.  Rescattering from exposed polarized iron in 
the spectrometer was estimated to be negligible. The net correction for 
all backgrounds was $0.6\pm 1.4\%$ for $^1$H and $2.8\pm 0.8\%$ for $\he$.

The measured asymmetry was corrected for the electron beam polarization 
using a Compton polarimeter with a precision of 1\%. The dominant 
systematic uncertainty in the Compton polarimetry arises from the 
determination of the analyzing power with realistic effects from detector 
resolution and energy calibration.  The key to high precision was the 
extremely precise beam tune, where the beam halo was so well contained at 
the Compton electron detector that the silicon strip detector could 
frequently make measurements with active strips only 5~mm from the primary 
electron beam.  This allowed a precision energy calibration of the 
electron detector, through measurement of the location of the 
cross section kinematic limit (the ``Compton edge'') and the zero-crossing 
of the asymmetry.  The energy calibration of the electron detector could 
then be transferred to the photon detector through spectra measured with 
an electron-photon coincidence trigger.  This technique reduced the 
calibration errors relative to previous analyses using the Compton 
polarimeter \cite{Baylac}, with the final uncertainty estimated at the 
level of 1\%.  Within the quoted uncertainties, the results were 
consistent (though somewhat more precise) with the measurements from the 
\Moller polarimeter in the hall and the Mott polarimeter in the injector.

\begin{figure}[phtb]\begin{center}
{\includegraphics[width=2.5in]{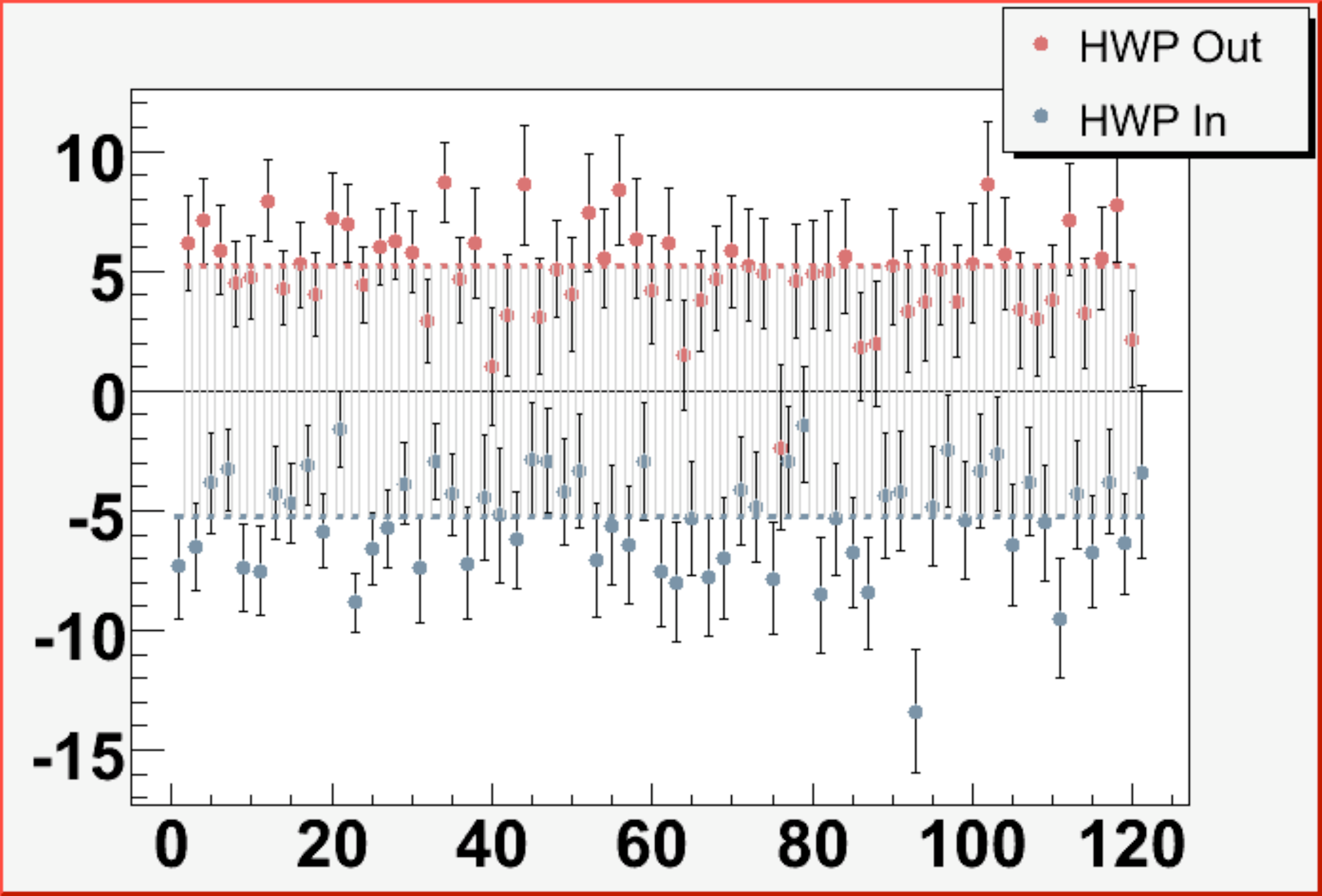},
\includegraphics[width=2.5in]{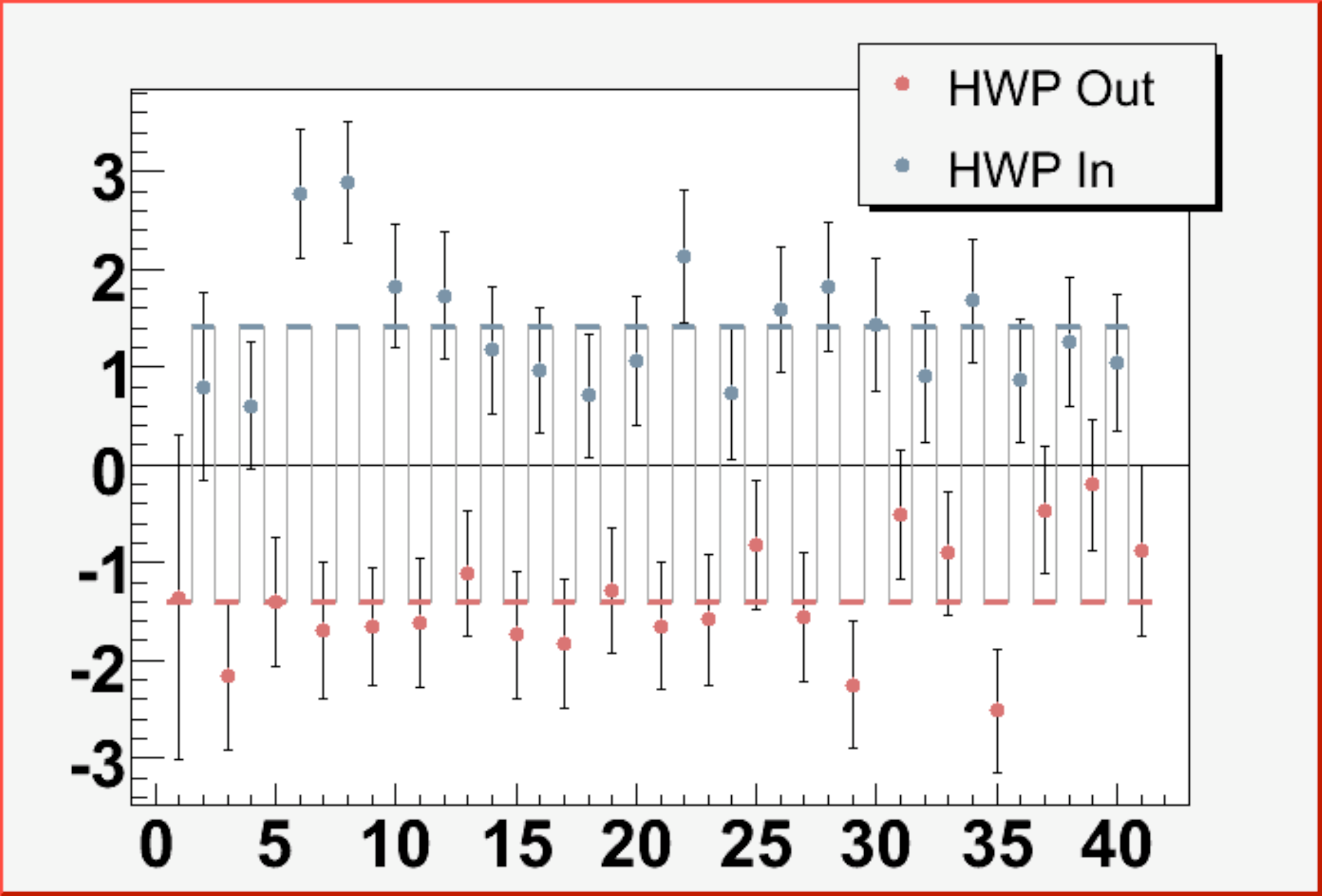}}
\caption{Corrected asymmetries (ppm) for HAPPEX-II in the 2005 run~
\protect\cite{Acha:2006my}.
Each data point represents the measured asymmetry, averaged over the two 
spectrometers, for a period between half-wave plate reversals, for the 
$\he$  (left) and $\hy$ (right) measurements. 
The step pattern represents the insertion and removal of the half-wave 
plate which reverses the sign of the asymmetry.  Vertical axis labels are 
in units of parts~per~million, horizontal axis is the index for periods 
between half-wave plate changes.}
\label{fig:happex_slug}
\end{center}\end{figure}

\begin{figure}[pbth]\begin{center}
\includegraphics[width=3.5in]{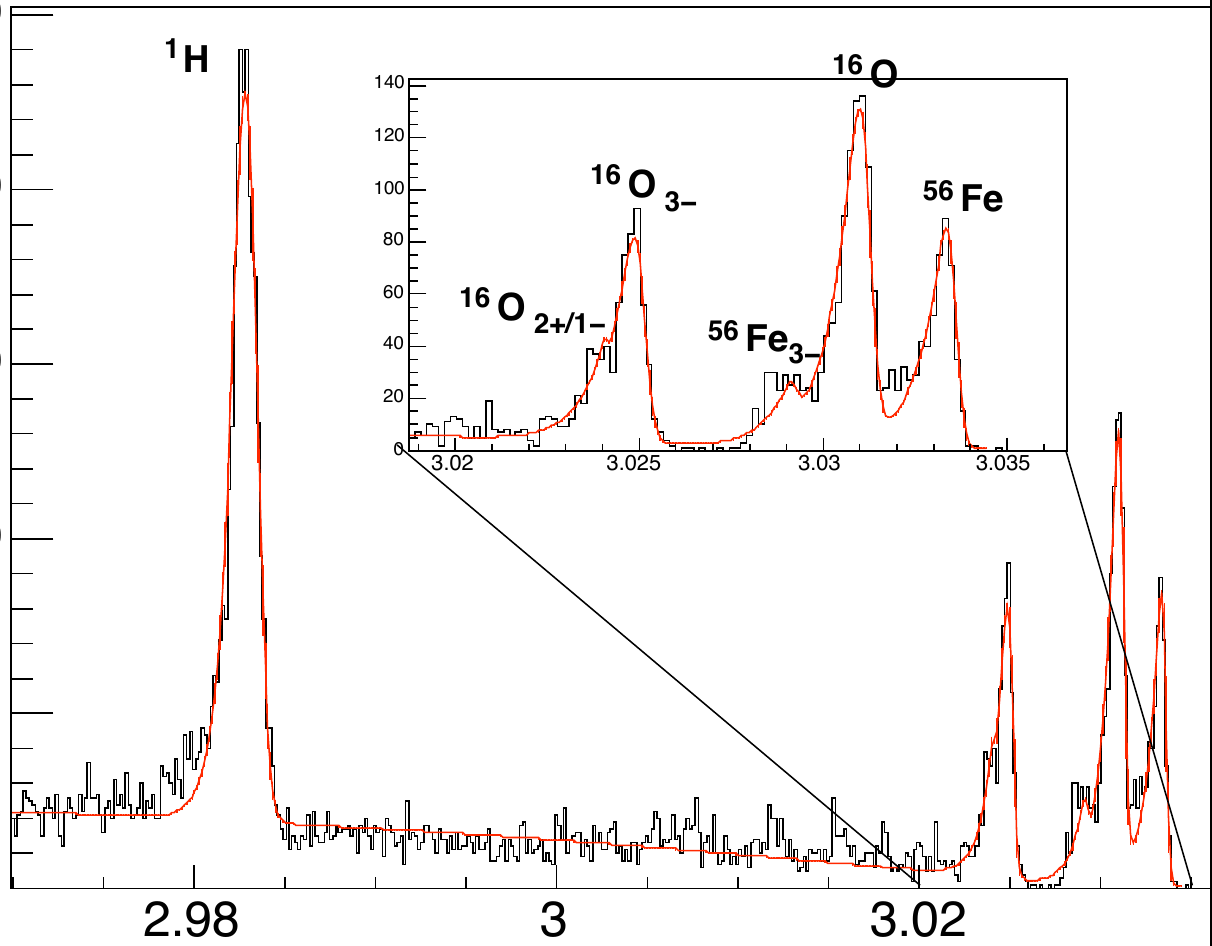}
\caption{Reconstructed momentum spectrum for the HAPPEX-II watercell 
target (horizontal axis in units of GeV).  Iron and oxygen states can be 
precisely resolved; measuring the momentum difference between these 
states and elastic proton scattering provides an accurate measurement of 
the scattering angle, with greatly reduced sensitivity to uncertainty in 
the primary beam momentum and spectrometer momentum calibration.  The 
availability of each state from a single target reduces the uncertainties 
in target energy loss corrections. Using this technique, scattering 
angles were calibrated with an accuracy of 0.2~mrad in an entirely beam 
based measurement.  Results agreed with survery reports on the 
experimental apparatus.
\protect\label{fig:happex_pointing} }
\end{center}\end{figure}

In order to interpret the measured asymmetry, an accurate determination of 
the accepted kinematics is required. A thin water target with steel beam 
window foil was used to determine the angular alignment of the 
spectrometer by observing the momentum difference between hydrogen elastic 
scattering and elastic or inelastic scattering from heavier nuclei, as 
shown in Fig.~\ref{fig:happex_pointing}.  Results from this calibration 
method agreed well with an independent determination based on survey of a 
pinhole collimator, which could be inserted between the target and any 
magnetic elements of the spectrometer.  As a result, uncertainties in the 
kinematics calibration were bounded to less than 1\%, leading to a 1.7\% 
systematic error of $^1$H and 0.9\% for $\he$.  An additional correction 
of 2.1\% $\pm$ 0.2\% for the hydrogen measurement, accounting for the 
finite range of $\qsq$ accepted, was determined using Monte Carlo 
simulation. The $\he$ measurement is exactly proportional to $Q^2$ at 
tree level and required no such correction.

\begin{table}\begin{center}
\begin{tabular}{|l|c|c|}
\hline
Correction (ppb) & Helium & Hydrogen \\ \hline
\hline
Beam Asymmetry & $183 \pm 50$ & $10 \pm 17$ \\ 
Target window bkg. & $ 113\pm 32$ & $ 7\pm 19$ \\ 
Helium QE bkg.& $ 12\pm 20$ & $-$ \\ 
Rescattering bkg.& $ 20\pm 15$ & $ 2 \pm 4$ \\ 
Nonlinearity& $ 0\pm 58 $ & $ 0 \pm 15 $ \\ \hline \hline
Scale Factor & Helium & Hydrogen \\ \hline \hline
Acceptance Scale Factor $K$& $ 1.000 \pm 0.001$ & $ 0.979 \pm 0.002$ \\ 
$Q^2$ Scale & $1.000 \pm 0.009$ & $ 1.000\pm 0.017 $ \\ 
Polarization $P_b$& $0.844 \pm 0.008$ & $0.871 \pm 0.009$ \\ \hline
\end{tabular}
\caption{\label{tab:HapSyst}Corrections and systematic error summary for HAPPEX-II and HAPPEX-Helium~\protect\cite{Acha:2006my}.}
\end{center}\end{table}

The published results for the HAPPEX-II $\hy$ and $\he$ measurements are 
given in Table~\ref{tab:H2results}.  The extracted results for the strange 
form factors from the hydrogen measurements are necessarily dependent on 
the values of the electromagnetic form factors, which introduces an 
uncertainty into the theoretical expectation for the asymmetry in the 
absence of the strangeness contribution. This extraction is also very 
weakly dependent on the poorly know ``anapole moment'' radiative 
correction, which influences the axial term. The extraction shown here 
uses the Zhu {\it et al.}~\cite{Zhu:2000gn} estimate for this correction 
and corresponding uncertainty. Under the simple assumption that $G_E^s$ is 
proportional to $Q^2$ and $G_M^s$ is varying slowly enough in this region 
to be considered constant, the electric and magnetic strange vector 
form factors can be extracted from these results to be: $G_E^s = -0.005 
\pm 0.019$ and $G_M^s = 0.18 \pm 0.27$, with a correlation coefficient of 
-0.87.

\begin{table}\begin{center}
\newcommand\B{\rule[-1.2ex]{0pt}{17pt}}
\begin{tabular}{|c|c|l|l|}
\hline
dataset & $Q^2$ [GeV$^2$]\B & \multicolumn{2}{c|}{ Result} \\ \hline\hline
2004 $\he$ & 0.077 &$A^{\rm He}_{NVS} = +7.48~\mbox{ppm}$
  & $A^{\rm He}_{PV} = +6.72 \pm 0.84 \pm 0.21  \mbox{ppm}$ \B\\  \hline 
& & \multicolumn{2}{c|}{$G_E^s = -0.038 \pm 0.042 \pm 0.010$} \B\\ \hline 
2005 $\he$ & 0.091 &  $A^{\rm He}_{NVS} = +6.37~\mbox{ppm}$ 
  & $A^{\rm He}_{PV} = +6.40 \pm 0.23 \pm 0.12 \mbox{ppm}$\B \\  \hline 
& & \multicolumn{2}{c|}{ $G^s_E = 0.002 \pm 0.014 \pm 0.007$} \B\\ \hline
2004 $^1$H & 0.099 & $A^{\rm H}_{NVS} = -1.43 \pm 0.11~\mbox{ppm}$
  & $A^{\rm H}_{PV} = -1.14 \pm 0.24 \pm 0.06 \mbox{ppm}$ \B \\ \hline
& & \multicolumn{2}{c|}{$G^s_E + 0.08 G^s_M = 0.030 \pm 0.025 \pm 0.006 \pm 0.012$} \B \\ \hline
2005 $^1$H & 0.109 & $A^{\rm H}_{NVS} = -1.66 \pm 0.05~\mbox{ppm}$
  & $A^{\rm H}_{PV} = -1.58 \pm 0.12 \pm 0.04 \mbox{ppm}$  \B\\  \hline 
& & \multicolumn{2}{c|}{$G^s_E + 0.09 G^s_M = 0.007 \pm 0.011 \pm 0.004 \pm 0.005$} \B \\\hline
\hline
\end{tabular}\caption{\label{tab:H2results} Published results from the 
HAPPEX-II measurements on hydrogen and helium~\protect\cite{Acha:2006my}. 
Error bars are listed {in order} for statistical and systematic 
uncertainties, and for form factor uncertainties where appropriate.}
\end{center}\end{table}

\section{The G0 experiment}

The G0 experiment~\cite{g0_forward,g0_backward}
measured the forward proton asymmetries and backward asymmetries
for both the proton and deuteron to provide, within a single apparatus,
a complete set of measurements over a broad range of
kinematics from which the electric and
magnetic strangeness form factors, as well as the 
axial neutral weak current of the nucleon, could be
extracted. 

\subsection{G0 forward}

The forward-angle G0 experiment~\cite{g0_forward} ran in Hall~C in
2003 and 2004, using a novel 8-coil, superconducting toroidal magnet
to focus recoil protons from the elastic electron-proton scattering
onto one of eight arrays of plastic scintillator detectors located
outside the magnet.  A $40~\mu$A, 74\% polarized beam from a strained
GaAs polarized source was used, with 32~ns pulse timing, rather than
the standard 2~ns, to allow for precise time-of-flight measurements
which discriminated against various backgrounds.  The delivery of 
high average beam
current with this pulse timing required very high peak currents,
and therefore provided significant challenges in accelerator physics
due to the large space-charge effects in each bunch.  The acceptance
was about 0.9 steradians at recoil angles centered around
$70^{\circ}$.  With a beam energy of 3~GeV the acceptance corresponded
to a $\qsq$ range $0.12 \le \qsq \le 1.0 \gevc$.  Each of the eight
detector arrays consisted of 16 different detectors, the Focal Plane
Detectors (FPD$_i$). The detectors were each a two-layer sandwich of
scintillators, and each layer was read out by two phototubes.  Each
detector in the range distinguished a unique bin in $\qsq$, with the
exceptions of the range $0.44 \le \qsq \le 0.88\gevc$, which all lay
in FPD 15 and was segmented using time-of-flight, and FPD 16, 
which was located beyond the kinematic limit for elastic
scattering, and was used for monitoring backgrounds.  The $\qsq$
acceptance for each detector was calibrated to 1\% precision using
time-of-flight measurements between $\pi^+$ and elastic protons.

\begin{figure}\begin{center}
\includegraphics[width=4in]{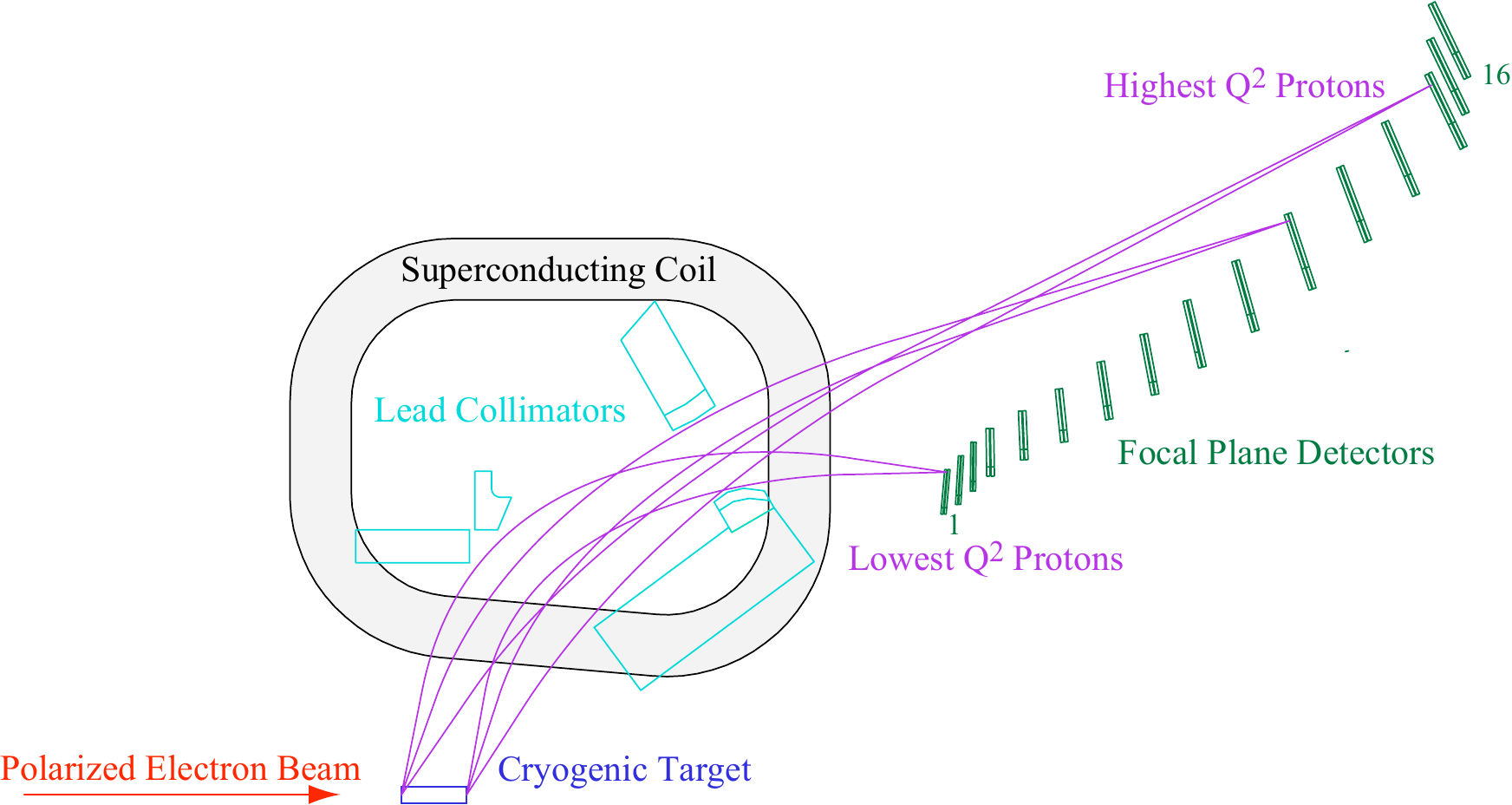}
\caption{Schematic of the G0 spectrometer as used in the forward angle 
experiment.}
\label{fig:G0spect}
\end{center}\end{figure}

\begin{figure}\begin{center}
\includegraphics[width=4in]{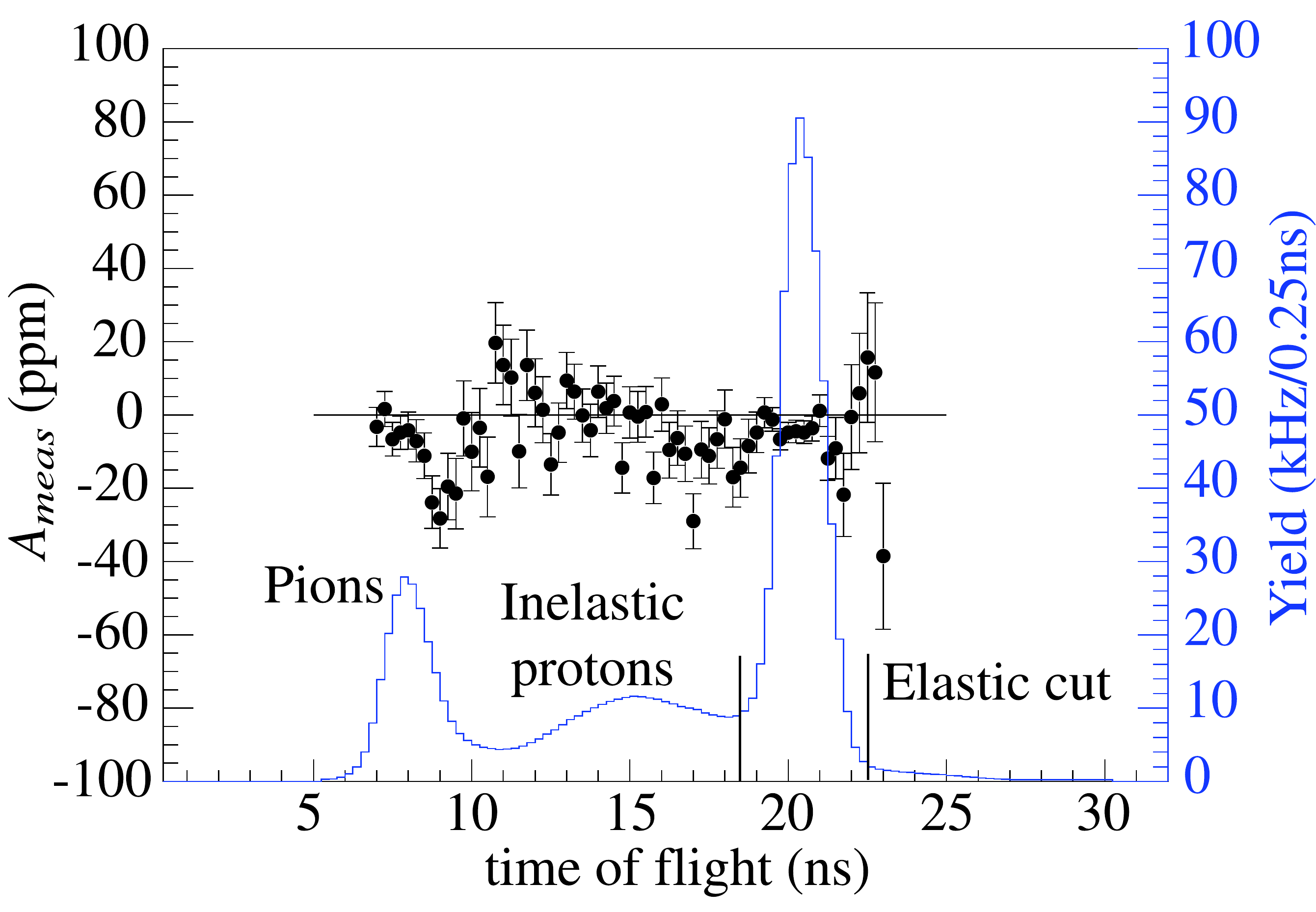}
\caption{The time-of-flight distribution (histogram) for a typical 
detector (FPD 8) in the forward angle G0 experiment, along with the
raw measured asymmetries (points with statistical errors).}
\label{fig:G0TOF}
\end{center}\end{figure}

Rather than an integrating method, as adopted by the HAPPEX experiments, 
the G0 experiment used a counting method to form the experimental 
asymmetries. Custom electronics allowed time-of-flight spectra to be 
accumulated for each helicity state. An asymmetry was then formed 
bin-by-bin in the spectra for each FPD. The pattern of fast helicity 
reversal was, unlike the HAPPEX experiments, formed not in pairs, but in 
quartets, where each quartet was either the sequence of helicity states 
LRRL or RLLR (L and R referring to left and right handed helicity, 
respectively). The quartet structure was chosen to exactly cancel the 
effect of any linear drifts in, for example, detector threshold or gain. 
The measured asymmetry was formed for each helicity quartet, {\it e.g.} 
$A_{\rm meas} = \frac{Y_L-Y_R-Y_R+Y_L}{Y_L+Y_R+Y_R+Y_L}$ for LRRL, where 
$Y_L$ and $Y_R$ are the measured yield in each helicity window.
This measured asymmetry can be expressed as
$A_{\rm meas}=(1 - f)A_{\rm el} + f A_{\rm b}$ where $A_{\rm el}$ is the 
raw elastic asymmetry, $A_{\rm b}$ the background asymmetry and $f$ the 
background fraction.

The elastic protons were identified by time-of-flight relative to the 
electron beam bunch; this allowed the rejection of protons from inelastic 
scattering, and faster particles such as the $\pi^+$, as shown in 
Fig.~\ref{fig:G0TOF}.  Background yields and asymmetries were measured 
concurrently and used to correct the elastic asymmetries.  In the region 
of the elastic peak, the background is essentially composed only of 
inelastic protons.  Background was subtracted in each detector using a 
simultaneous fit of the time-of-flight yield and asymmetry spectra near 
the elastic cut.  For this fit, the yield is modeled with a Gaussian 
elastic peak on a polynomial background, while the elastic asymmetry is 
fit to a constant and the background to a quadratic function.
For higher number detectors the background asymmetry becomes positive, 
due to a small number of protons from the weak decays of hyperons, that 
scatter in the magnet, leading to corrections which were 20-110\% of the 
final result. The uncertainties in these corrections were estimated from 
the variation in results using a variety of different models for the 
background asymmetries and yields.   These corrections lead to a 
significant correlation between the estimated systematic uncertainties 
for data points in the range $\qsq > 0.3\gevc$. 

The beam polarization was measured using a M{\o}ller 
polarimeter~\cite{Hauger:2001} to a precision of 1\%.  Radiative 
corrections of 1--3\% were performed using simulation. Small corrections 
for electronics deadtime were made on the basis of the observed dependence 
of the yield on the beam current. An additional correction was required 
for an asymmetric ``leakage'' beam.  This was a problem unique to the 
non-standard time structure: some fraction of the total beam to Hall~C 
was leakage beam, primarily tails of the beam intended for the other two 
halls, which arrived with the standard 499~MHz time structure and was 
recorded by the 1497~MHz beam current monitors along with the expected G0 
beam.  Being off-time, elastic scattered protons from this leakage beam 
did not contribute to the signal in the elastic cut, leading to a 
mis-match between the measured beam current asymmetry and the beam current 
which contributed to the accepted signal. Although a small fraction of the 
total beam, there was a strong helicity dependence for this leakage beam.  
The correction for this leakage beam asymmetry was $0.7 \pm 0.1$~ppm, and 
nearly uniform over all detectors.

\begin{figure}\begin{center}
\includegraphics[width=4.5in]{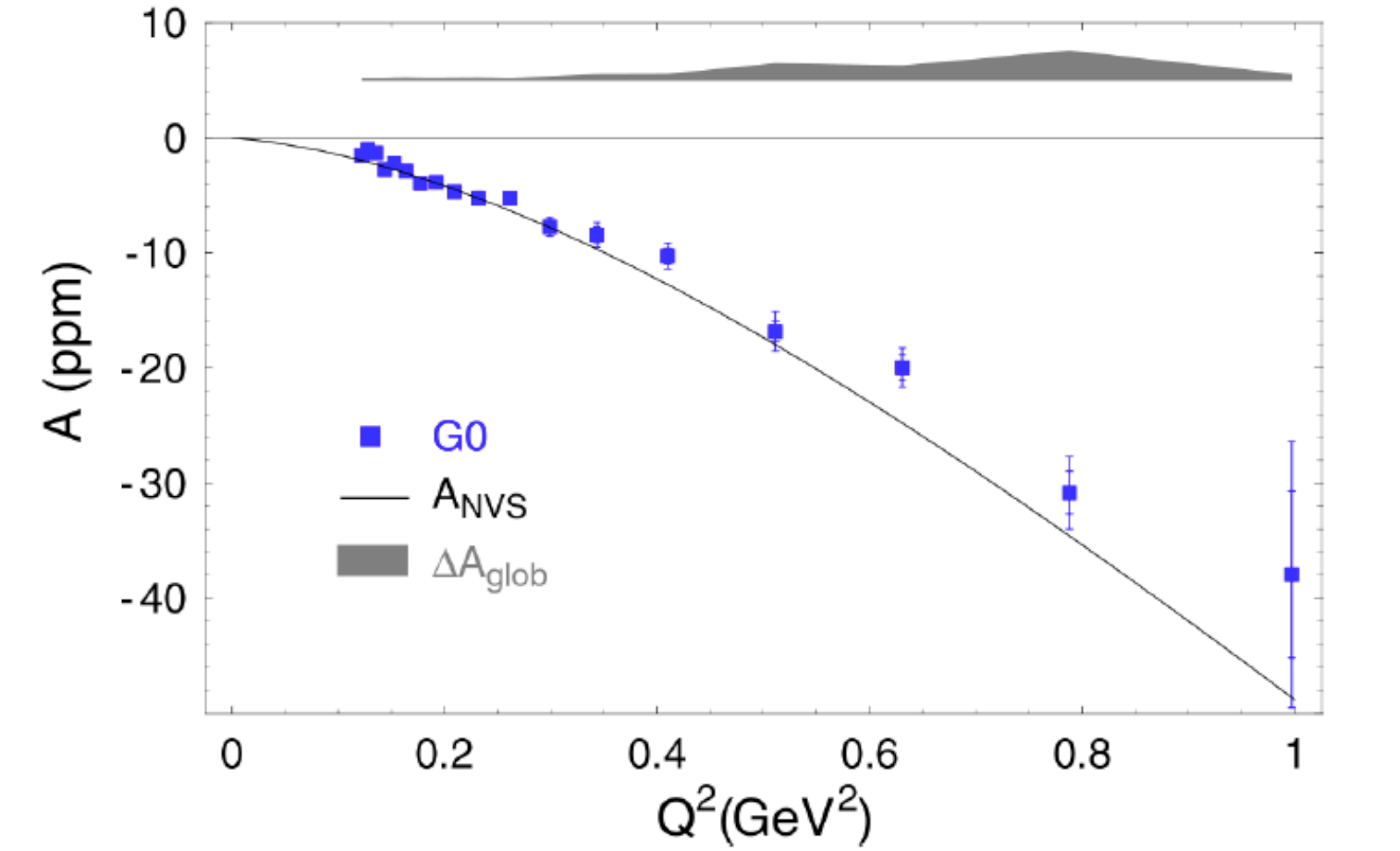}
\caption{Results from the G0 forward angle measurements of $\apv$ in 
elastic electron-proton scattering~\protect\cite{g0_forward}.  Data points 
are shown with statistical and point-to-point systematic error bars 
combined, while the shaded band represents the magnitude of systematic 
errors which are highly correlated between the data points.  The curve 
A$_{\rm NVS}$ represents the expected value of $\apv$ in the absence of 
strange contributions to the vector form factors.}
\label{fig:G0AsymQ2}
\end{center}\end{figure}

Results for the final measured asymmetries after all corrections are shown 
as a function of $\qsq$ in Fig.~\ref{fig:G0AsymQ2}. For each $\qsq$ point, 
the combination of strangeness form factors $G^s_E + \eta G^s_M$ as a 
function of momentum transfer was obtained, as discussed in 
Section~\ref{sec:pvesSummary} and shown in Fig.~\ref{fig:SummaryVsQ2}, 
where $\eta(Q^2) = \tau G^p_M / \epsilon G^p_E$.  The results were in 
excellent agreement with the earlier HAPPEX-I 
measurement~\cite{happex1_prc}, and displayed a noticeable bias toward a 
positive strangeness contribution at higher $Q^2$, which prompted the 
next round of high-precision experiments at Jefferson Lab and at Mainz.

\subsection{G0 backward}

\begin{figure}\begin{center}
\includegraphics[width=5in]{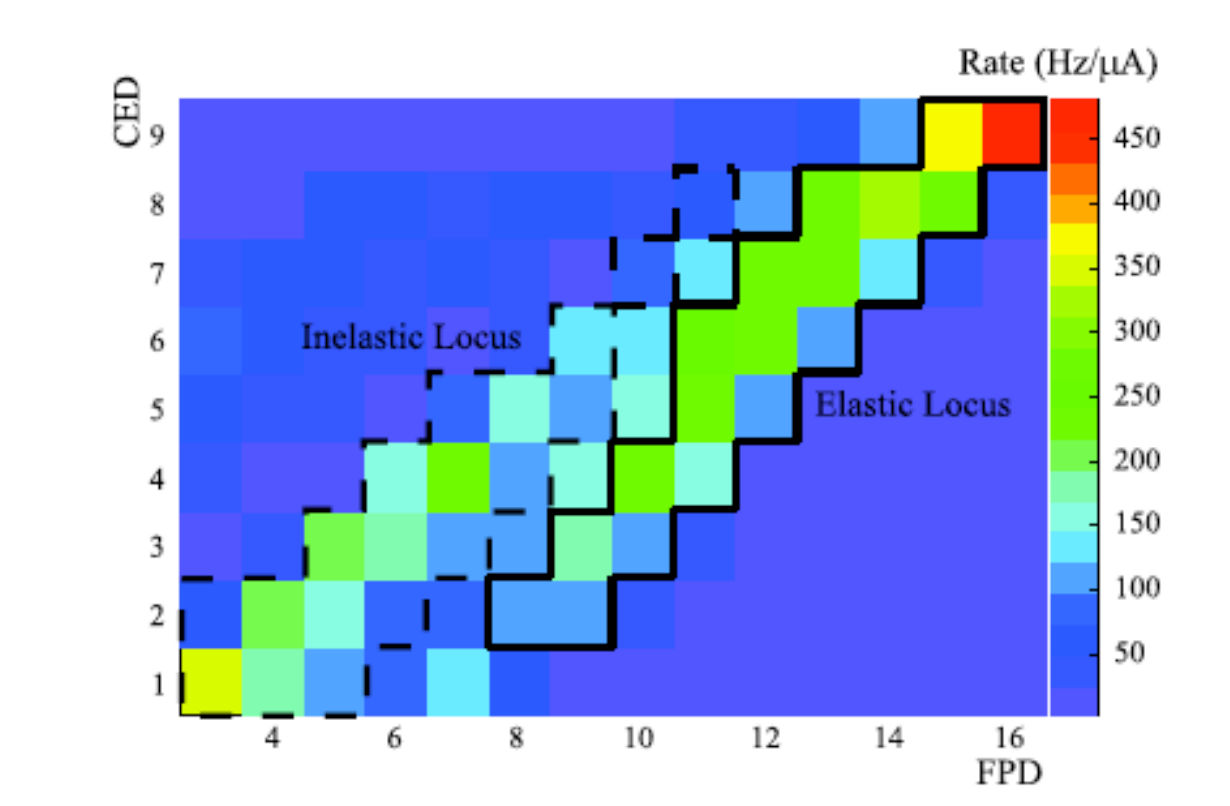}
\caption{The distribution of measured yields from the backward-angle
  phase of G0, as a function of FPD and CED, for hydrogen data taken
  at a beam energy of 687~MeV.  (Note: FPDs 1 and 2 are not used,)
  Electrons from elastic (inelastic) scattering are in the upper
  right (lower left).}
\label{fig:G0Matrix}
\end{center}\end{figure}

The G0 backward-angle phase experiment was conducted in Hall~C at
Jefferson Lab in 2006 and 2007.  The G0 superconducting toroidal spectrometer
was turned 180$^{\circ}$ around with respect to the beam direction compared
to the forward-angle configuration, in order to detect
the electrons scattered at an angle of about $110^{\circ}$ from 20~cm
liquid hydrogen and deuterium targets. 
Polarized electron beams with currents up to 60~$\mu$A
and energies of 359 and 684~MeV were generated with a strained GaAs
polarized source. In this phase, the standard Jefferson Lab time 
structure was used for the beam, as time-of-flight was not useful for 
separating the scattered electrons from backgrounds. The quartet 
helicity reversal pattern was adopted as for the forward-angle 
experiment. 
The average beam polarization was $85.8 \pm  2.1(1.4)\%$ at the lower 
(higher) incident energy. In this phase, the focal-plane detector array 
(FPD$_i$) that was used in the forward angle measurement was augmented 
with a second array of scintillators, near the exit of the magnet 
(CED$_i$).
Coincidences between these two scintillator arrays allowed electrons 
from elastic and inelastic scattering to be distinguished, as shown
in Figure~\ref{fig:G0Matrix}.

Additionally, in each octant  an aerogel Cerenkov detector was added
in order to separate the electrons from the copious background from
pions. The Cerenkov detectors had a pion rejection factor
$\geq 85\%$ and an electron efficiency of about 85\%.
The backgrounds in the region of the elastic locus (see Fig.~\ref{fig:G0Matrix}) 
amounted to 10--15\% of the signal. In the elastic locus, the aluminum
target windows was the dominant background with misidentified $\pi^-$ 
and electrons from $\pi^0$ conversion also contributing.
The aluminum fraction was measured using runs with cold gaseous 
hydrogen in the target and the pion contamination was determined
from dedicated time-of-flight runs and pulse-shape analysis. 
Additionally, an acceptance study was performed by sweeping the field 
of the toroid $\pm 40\%$ of the nominal setting and comparing the 
measured yield in each coincidence cell with Monte Carlo simulation 
using GEANT.
The aluminum asymmetry was taken to be the same as that of the deuteron 
(both effectively quasi-elastic scattering only) with an additional 
uncertainty of 5\% for nuclear effects. The asymmetry for the pion 
production background was small ($< 1$~ppm) and was measured concurrently 
with that for the electrons.
The background corrections are small because the background asymmetries 
generally have values close to those of the elastic asymmetry, or, 
otherwise, the fraction is small.

All asymmetries were corrected for measured rate-dependent effects. For 
elastic scattering, dead-time corrections generally dominated those from 
accidentals and amounted to $\sim15$\% of the yield based on the measured 
beam current dependence, and led to an uncertainty of about 0.5~ppm in the 
asymmetries.
In the high-energy deuteron measurement, accidentals from pion signals in 
the scintillators in coincidence with random signals from the Cerenkov 
dominated the correction. In this case, the correction to the asymmetry 
was $7.0 \pm 1.8$~ppm.
Helicity-correlated intensity changes were corrected with active feedback 
to about 0.3~ppm.  Corrections to the measured asymmetry for residual 
helicity-correlated beam current, position, angle and energy variations 
of at most $0.2 \pm 0.07$~ppm were applied {\it via} linear regression.
Electromagnetic radiative corrections of $(3--3.5) \pm 0.3$\% and small 
two boson exchange effects (1\%) were also applied to the asymmetries. 
Table~\ref{Tab:G0BackCor} shows the corrections to the raw elastic 
asymmetry, $A_{\rm el}$, as well as the final asymmetries $A_{\rm phys}$ 
and their statistical and systematic uncertainties. The extraction of 
strange form factors from these data is discussed in the next section.

\begin{table}\begin{center}
\caption{\label{Tab:G0BackCor} G0 backward-angle: corrections to the 
 measured asymmetry and the final physics asymmetries.
 Rate and ``Other'' corrections are additive; 
 electromagnetic
 radiative corrections are multiplicative.  ``Other'' corrections
 include those for helicity-correlated beam parameters, the
 transverse component of beam polarization, and two-boson exchange.
 Not listed is the common multiplicative correction for the 
 beam polarization, $1.16\pm 0.02$.
 The uncertainties for the asymmetries are statistical, point-to-point 
 systematic and global systematic, respectively. }
\begin{tabular}{ccccccc}
Target & $Q^2$ & Rate & Other & EM Radiative & $A_{\rm phys}$\\
&(GeV$^2$)& (ppm) &(ppm) & & (ppm) \\
\hline
H & 0.221 & $-0.31 \pm 0.08 $ & $0.22 \pm 0.08$  &  $1.037 \pm 0.002 $ & $-11.25 \pm 0.86 \pm 0.27 \pm 0.43$\\
D & 0.221 & $-0.58 \pm 0.21 $ & $0.06 \pm 0.10 $ &  $1.032 \pm 0.004$ & $-16.93 \pm 0.81 \pm 0.41 \pm 0.21$\\
H & 0.628 & $-1.28 \pm 0.18 $ & $0.29 \pm 0.11 $ &  $1.037 \pm 0.002$ & $-45.9 \pm 2.4 \pm 0.8 \pm 1.0$\\
D & 0.628 & $-7.0 \pm 1.8 $ & $0.34 \pm 0.21 $   & $1.034 \pm 0.004$ & $-55.5 \pm 3.3 \pm 2.0 \pm 0.7$\\
\end{tabular}
\end{center}\end{table}

\section{Summary and outlook}
\label{sec:pvesSummary}

\begin{figure}\begin{center}
\includegraphics[width=5in]{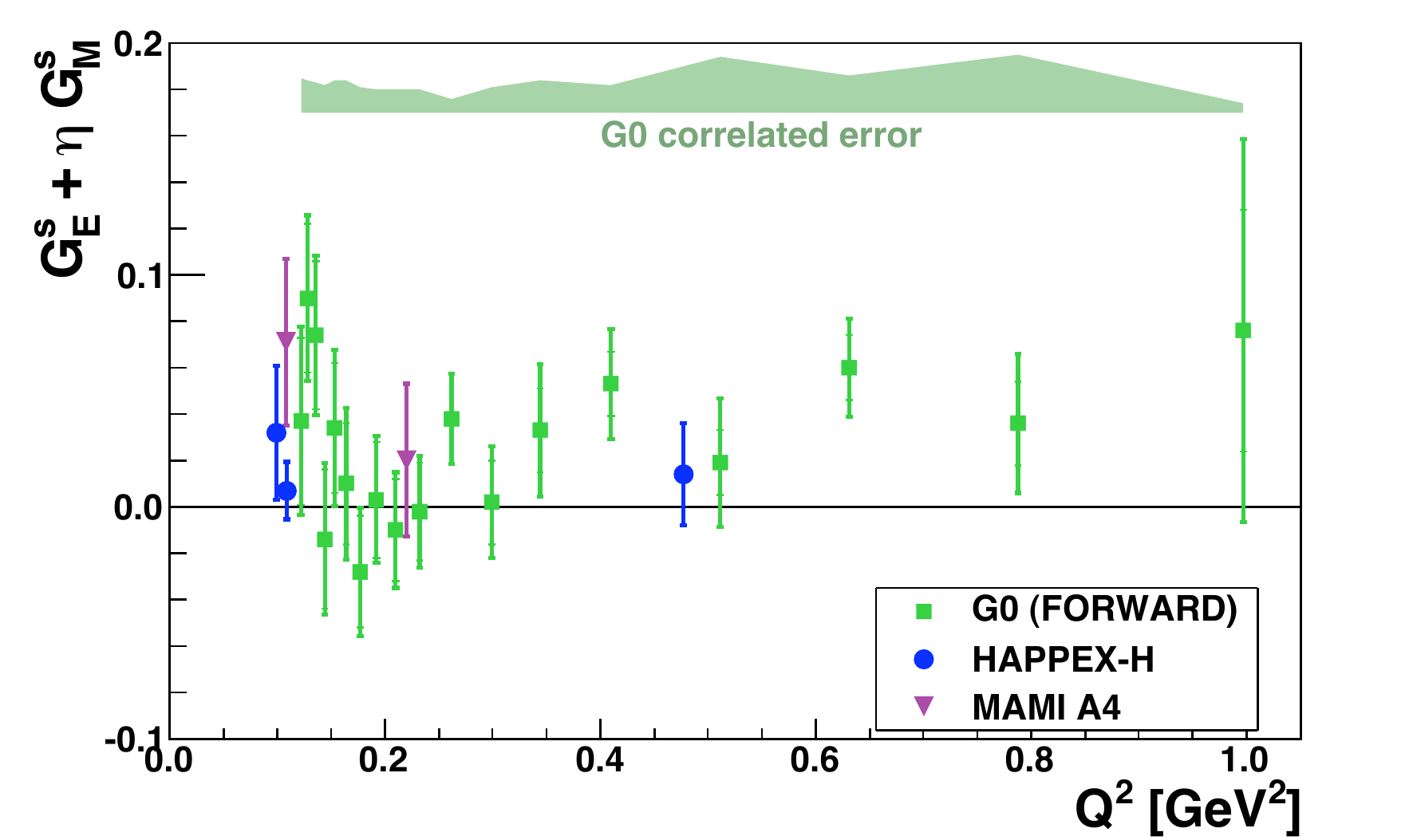}
\caption{Results from world data measuring $A_{PV}$ from forward-angle scattering from the proton, plotted as the net 
strange quark contribution $G_E^s + \eta G_M^s$.  Here $\eta = \tau G_E^p/{\epsilon G_M^p}$, and is approximately equal to $\qsq$ over this plot. In addition to HAPPEX and G0, published
results from the PVA4 experiment at MAMI at $\qsq = 0.1 \mbox{ and } 0.23\gevc$\ are shown
\protect\cite{Maas:2004ta,Maas:2004dh}.\protect\label{fig:SummaryVsQ2}}
\end{center}\end{figure}

The HAPPEX and G0 experiments at Jefferson Lab have provided precision 
measurements of the weak form factors of the nucleon over a range of 
$\qsq < 1\gevc$.  The global data set on forward-angle scattering from 
the proton, which is shown in Fig.~\ref{fig:SummaryVsQ2}, also includes 
measurements from the PVA4 collaboration at the Mainz Microtron 
\cite{Maas:2004ta,Maas:2004dh}.  For each data point, the combination of 
strangeness form factors $G^s_E + \eta G^s_M$ as a function of momentum 
transfer was obtained, where $\eta(\qsq) = \tau G^p_M / \epsilon G^p_E$. 
This combination represents the net contribution of the strange form 
factors to $\alr$ in each measurement. Taken as a whole, this data set 
systematically appears to suggest a small but positive strange form factor 
combination, however, it does not establish a clear, statistically 
significant signal.  At low $\qsq$, the highest precision measurements 
constrain contributions to be very near zero.  At higher $\qsq>0.5\gevc$, 
there remains a suggestion of a possibly measurable contribution.

\begin{figure}\begin{center}
\includegraphics[width=5in]{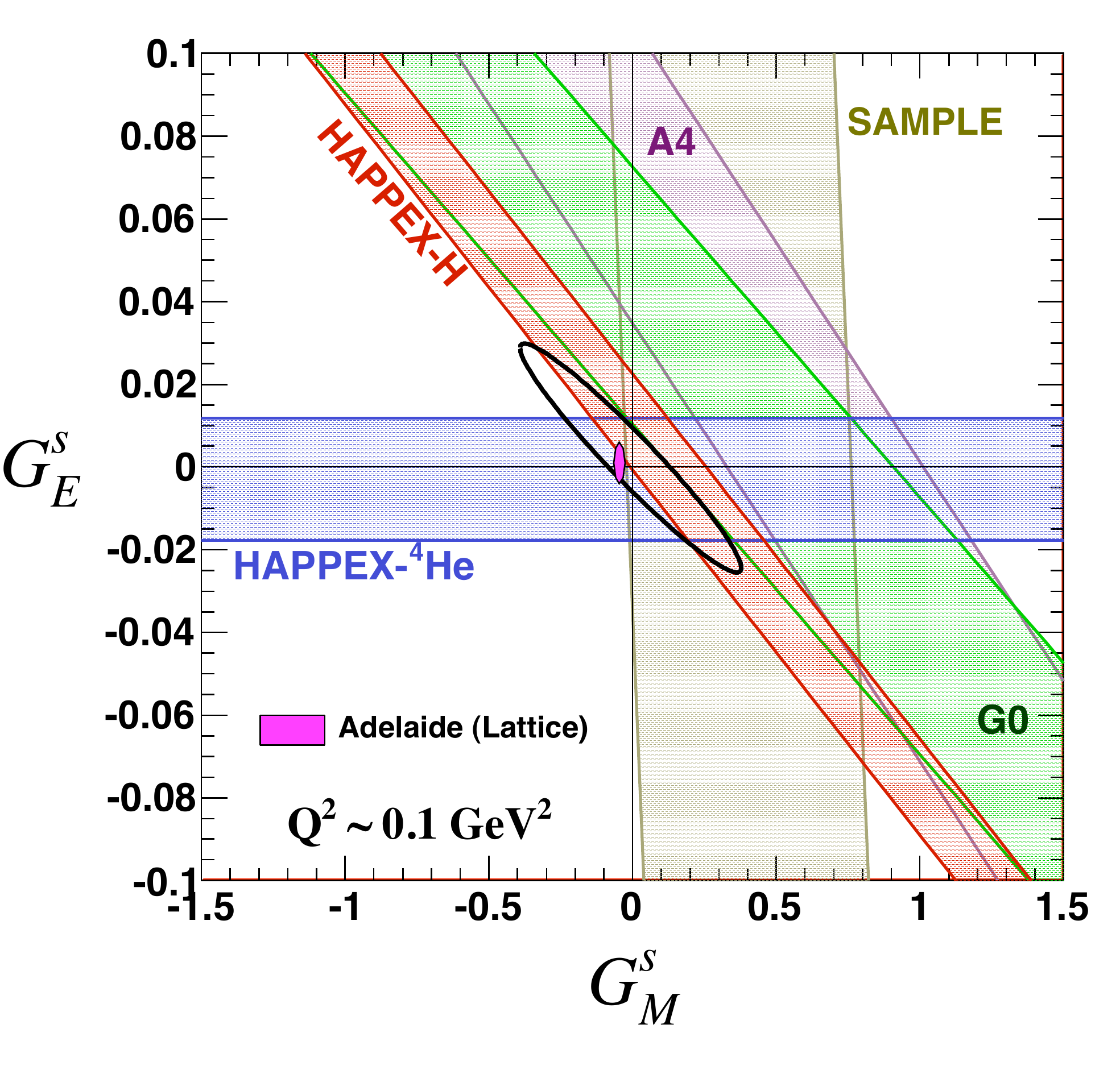}
\caption{Measurements of $\ges$ adn $\gms$ in the region of 
$\qsq=0.1\gevc$.  Each band represents the 1$\sigma$ error bar for a 
unique combination of the strange form factors.  Results of a recent QCD 
calculation from the Adelaide 
\protect\cite{Leinweber:2004tc,Leinweber:2006ug,Wang:1900ta} is shown 
as a filled ellipse. As described in the text, the 68\% confidence level 
contour of a global fit~\protect\cite{Young:2006jc} incorporating data 
beyond what is shown in this plot is shown as a solid elliptical 
contour.\protect\label{fig:Summary0.1}}
\end{center}\end{figure}

Figure~\ref{fig:Summary0.1} shows published data obtained at 
$\qsq\sim0.1\gevc$. In addition to forward-angle data from Jefferson Lab 
and PVA4~\cite{Maas:2004ta,Maas:2004dh} collaborations, this figure 
displays results from the HAPPEX $\he$ measurement and backward-angle 
measurements from SAMPLE at MIT-Bates~\cite{SAMPLE}. Each measurement 
appears as a band, representing the central value and 1$\sigma$ error bar 
for the specific combination of $\ges$ and $\gms$.  The G0 result plotted 
here is an average of 3 published data points from the forward-angle data 
set, covering $\qsq=0.12$ -- $0.14\gevc$. For each of these measurements 
the anapole radiative correction to the axial form factor is included with 
the uncertainty as prescribed by theoretical 
expectation~\cite{Zhu:2000gn}.  These data can be fit to extract
$G_E^s = -0.006\pm 0.016$ and $G_M^s = 0.33 \pm 0.21$, with a correlation 
coefficient of $-0.83$~\cite{Liu:2007}.  Also displayed, as the small, 
filled ellipse, is the result of the Adelaide indirect lattice calculation 
described in Section~\ref{sec:pvestheory:indirect}.

In the absence of reliable theoretical guidance on the $\qsq$ variation of 
the strange form factors, fits to the global data segt have used a 
first-order power-expansion in $\qsq$ to fit all data up to about $\qsq < 
0.3\gevc$.  The result of such a fit~\cite{Young:2006jc} is shown as the 
solid elliptical contour (68\% confidence level) in 
Fig.~\ref{fig:Summary0.1}.  This fit does not include the backward-angle 
results at $\qsq\sim0.22\gevc$ from G0 and A4, and applies no theoretical 
constraint on the axial form factor contribution which includes the 
anapole moment correction.  These results are statistically consistent 
with fits of the data near $\qsq=0.1\gevc$ which incorporate theoretical 
limits on the anapole correction, as shown above.  However, the inclusion 
of data at higher $\qsq$ tends to draw the central value of the fit to 
zero, implying either a strong $\qsq$ dependence to the strange form 
factors or that the suggested deviation at $\qsq\sim0.15\gevc$ is a 
statistical fluctuation.

\begin{figure}\begin{center}
\includegraphics[height=5in]{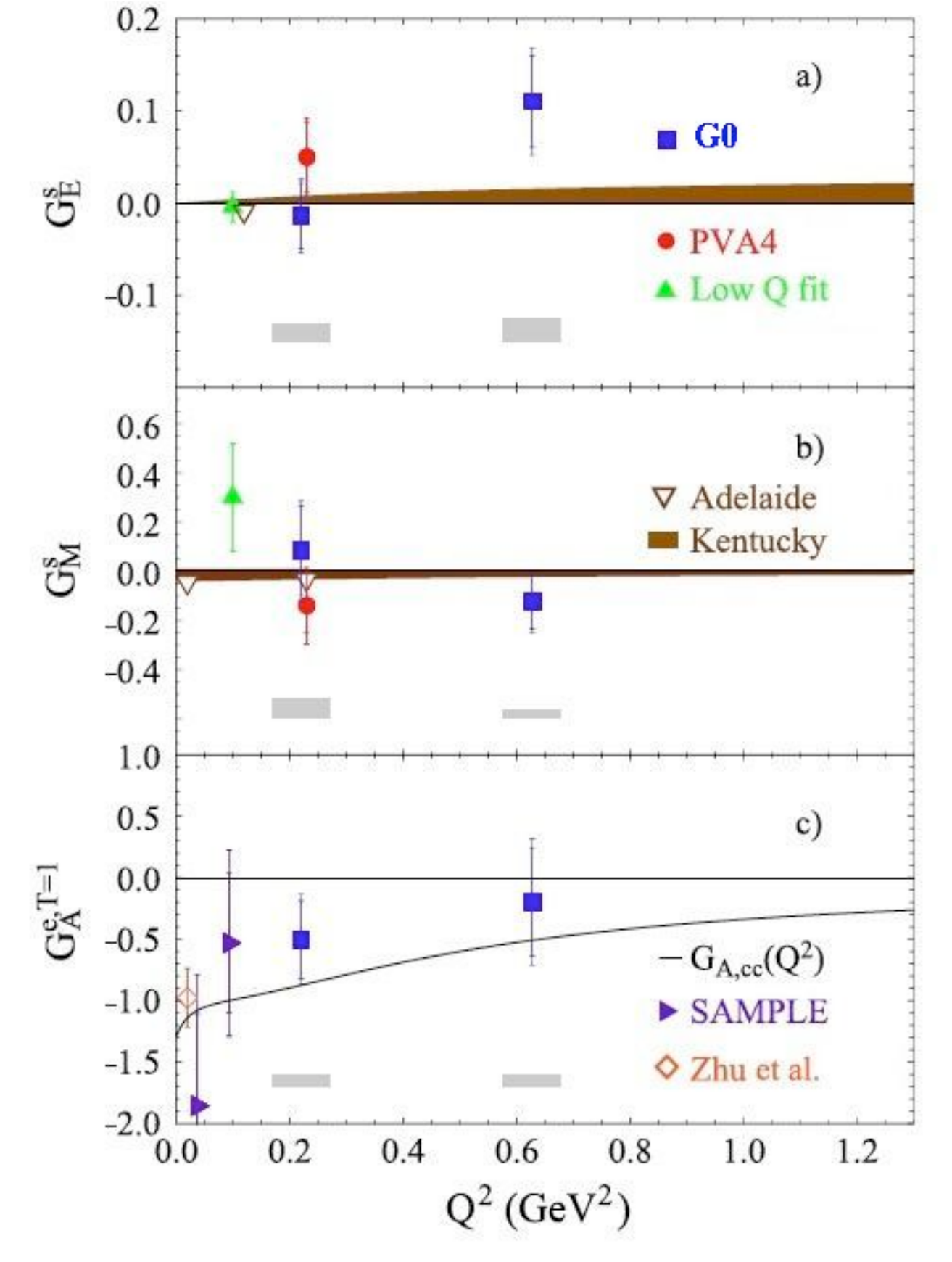}
\caption{The form factors a) $G^s_E$, b) $G^s_M$, and c) $G^e_A(T=1)$ 
determined by combination of the G0 experiment forward and backward-angle 
measurements. Error bars show statistical and statistical plus 
point-to-point systematic uncertainties (added in quadrature); shaded bars 
below the corresponding points show global systematic uncertainties (for 
the G0 points). For $G^s_E$ and $G^s_M$, the extraction from 
\protect\cite{Liu:2007} as well as the results of the PVA4 experiment 
\protect\cite{Baunack:2009} are shown. Recent QCD calculations from groups 
at Adelaide \protect\cite{Leinweber:2004tc,Leinweber:2006ug,Wang:1900ta} 
and Kentucky \protect\cite{Doi:2009sq} are also shown; for the former the 
uncertainties are smaller than the symbols. For $G^e_A(T=1)$, results from 
the SAMPLE experiment~\protect\cite{SAMPLE} are shown together with the 
calculation of Zhu {\em et al.}~\protect\cite{Zhu:2000gn}.}
\label{fig:G0Backward}
\end{center}
\end{figure}

Figure~\ref{fig:G0Backward} shows the strange form factors $G^s_E$,
$G^s_M$, and the isovector axial form factor $G^e_A(T=1)$, extracted 
from the combination of the G0 backward angle and forward-angle 
asymmetries at $\qsq=0.221$ and $0.628\gevc$~\cite{g0_forward,g0_backward}.
 These results utilize a simple
interpolation of the G0 forward angle measurements to the exact $\qsq$
of the backward angle data. The Kelly \cite{Kelly} parameterization of 
the nucleon electromagnetic form factors, $G^{p,n}_{E,M}$
was adopted for these determinations in order to be consistent with the 
deuteron model used~\cite{Schiavilla:2004}. Figure~\ref{fig:G0Backward} 
also shows the extraction of $G^s_E$, $G^s_M$,  $\qsq = 0.1~\gevc$ 
described above~\cite{Liu:2007}. 
Note that the PVA4 points shown~\cite{Baunack:2009}, in contrast to the 
G0 results, are not based on measurements on deuterium and therefore 
assume a value for $G^e_A(T=1)$ determined by theoretical expectation 
Zhu {\em et al.}~\cite{Zhu:2000gn} shown in Fig.~\ref{fig:G0Backward}c), and a dipole form 
factor with a mass parameter of 1.032~GeV.

Taken as a whole, all these data clearly suggest that the strange vector 
form factors are a small contribution to the charge and magnetization 
distributions of the proton, consistent with the recent direct and 
indirect lattice QCD determinations, discussed above.  There remains a 
suggestion in the experimental data that a significant contribution may be 
measurable at $\qsq\sim0.6\gevc$.  This suggestion will be tested by the 
HAPPEX-3 experiment~\cite{HAPPEX3}, which ran in 2009, measuring at a 
$\qsq\sim0.62\gevc$ and using techniques developed during the preceding 
HAPPEX experiments.  Benefiting from the high luminosity and polarization 
available at Jefferson Lab, along with improved alignment and polarimetry 
techniques, HAPPEX-3 will measure $G_E^s + 0.52 G_M^s$ with an error bar 
$\pm 0.010~\mbox{(stat)} \pm 0.005~\mbox{(syst)}\pm 0.010~\mbox{(ff)}$, 
where the latter uncertainty arises from imperfect knowledge of the 
electromagnetic form factors. This result will greatly improve the 
precision of the $\ges$, $\gms$ extraction at this $\qsq$, providing an 
opportunity to establish a clear, statistically significant non-zero 
signal.

This program of measurement was originally motivated by theoretical 
suggestions of potentially large contributions of strange quarks to the 
nucleon vector form factors. Over the kinematic range that has been 
examined, contributions have been found to be small or consistent with 
zero, with the most precise measurements having now pushed near to the 
bounds of unique interpretability.  One significant limit is the precision 
in the knowledge of electromagnetic form factors. Presently, uncertainties 
in the neutron form factors $G_E^n$ and $G_M^n$ contribute an uncertainty 
to the extraction of the strange form factors of about half of the 
statistical error bar of the most precise $\apv$ measurements. There are 
important radiative corrections as well.  As discussed above, the anapole 
moment correction in the axial term is difficult to bound with a precision 
approaching that of published theoretical estimates~\cite{Zhu:2000gn}. The 
precision of extraction of $\ges$ and $\gms$ is dependent on the 
reliability of calculation of this correction. Similarly, $\gamma Z$ and 
$\gamma\gamma$ box diagrams present a potentially important correction to 
the interpretation of measurements of $\alr$. Recent theoretical 
results~\cite{Tjon:2007wx,Gorchtein:2008px,Tjon:2009hf,Zhou:2009nf,Sibirtsev:2010zg} 
indicate that these uncertainties are sufficiently controlled for the 
current program, but significant advances beyond the next round of 
measurements would likely require further investigation into these 
corrections.

The extraction of the strange vector form factors also depends on the 
assumption of charge symmetry in the nucleon form factors.  While this was 
once thought to provide a negligible uncertainty~\cite{ChargeSymm}, recent 
work in has shown that charge symmetry violation in the nucleon may 
contribute to the form factor at a level comparable to the statistical 
error of the most precise current 
measurements~\cite{Viviani:2007,Kubis:2006cy}. For this reason, future 
measurements might best be interpreted more generally, as probing all 
mechanisms for charge symmetry breaking in the nucleon and not primarily 
the strange quark content.

As described later in this volume~\cite{BSM}, the studies of low-$\qsq$ 
elastic proton electroweak scattering from the strange quark program can 
be combined to provide a useful constraint on parameters of the Standard 
Model.  The Qweak experiment~\cite{Qweak} at Jefferson Lab will take 
advantage of this in a future measurement of $\alr$ from proton 
scattering at very low $\qsq$, in a sensitive new search for physics 
beyond the Standard Model.
Qweak aims for a precision which is more than an order of magnitude 
beyond any previous Jefferson Lab measurement. This underlines the fact 
that the HAPPEX and G0 experiments at Jefferson Lab have been successful 
not only in the measurement of strange form factors, but have forged a 
path for new measurements of parity-violation in electron scattering.
These significant advances in production and control of polarized beam, 
electron beam polarimetry, high luminosity targets, low noise integration 
techniques, and electron beam monitoring have enabled a new generation of 
parity violation experiments.  In this way, the legacy of the strange 
quark program extends to Qweak;  to a very challenging measurement of the 
neutron radius in $^{208}$Pb (PREX) \cite{PREX} which is so crucial in 
models of neutron-rich nuclei;  to a measurement of parity-violation in 
deep inelastic scattering ~\cite{PVDIS} which will improve precision on 
the poorly measured electroweak axial quark charges ($C_{2q}$); and to 
the measurements (described in~\cite{BSM}) of electroweak couplings and 
valance parton distributions planned after the upcoming energy upgrade 
at Jefferson Lab~\cite{MOLLER,SOLID}.

\end{document}